\documentclass[lettersize, journal]{IEEEtran}
\usepackage{amsmath,amsfonts}
\usepackage{algorithmic}
\usepackage{array}
\usepackage[caption=false,font=normalsize,labelfont=sf,textfont=sf]{subfig}
\usepackage{textcomp}
\usepackage{stfloats}
\usepackage{url}
\usepackage{verbatim}
\usepackage{graphicx}
\usepackage{tabularx}
\usepackage{authblk}
\def\BibTeX{{\rm B\kern-.05em{\sc i\kern-.025em b}\kern-.08em
    T\kern-.1667em\lower.7ex\hbox{E}\kern-.125emX}}
\usepackage{balance}
\usepackage{enumitem}

\usepackage{color}
\newif\ifacm
\acmtrue

\newif\ifcomment
\commenttrue

\newif\ifcameraready
\camerareadyfalse

\newif\ifwatermark
\watermarkfalse

\newif\ifdiff
\difffalse

\ifdiff
    \newcommand{\del}[1]{{{\color{red}\st{#1}}}}
     
\else
    \newcommand\del[1]{}
    
\fi

\ifwatermark
      \usepackage{draftwatermark}
      \SetWatermarkScale{.7}
      \SetWatermarkText{\shortstack{In Submission:\\Do Not Distribute}}
      \SetWatermarkColor[rgb]{1,0.88,0.88}
\fi

\ifcomment
    \newcounter{MVNumberOfComments}
    \stepcounter{MVNumberOfComments}
    \newcommand{\mvnote}[1]{\textcolor{blue}{\small \bf [MV\#\arabic{MVNumberOfComments}\stepcounter{MVNumberOfComments}: #1]}}

    \newcounter{DJNumberOfComments}
    \stepcounter{DJNumberOfComments}
    \newcommand{\djnote}[1]{\textcolor{red}{\small \bf [DJ\#\arabic{DJNumberOfComments}\stepcounter{DJNumberOfComments}: #1]}}

    \newcounter{YZNumberOfComments}
    \stepcounter{YZNumberOfComments}

    \newcommand{\NOTE}[1]
    {
      {\footnotesize\it
        \begin{center}
          \begin{tabular}{|c|}
           \hline
            \parbox{0.85\columnwidth}{
              \medskip
              #1
              \medskip} \\
            \hline
          \end{tabular}
        \end{center}
        }
    }
\else
    \newcommand\mvnote[1]{}
    \newcommand\djnote[1]{}
    \newcommand\NOTE[1]{}
\fi

\newcounter{NumTakeaways}
\stepcounter{NumTakeaways}

\begin{document}

\title{A Tale of Three Location Trackers: \\ AirTag, SmartTag, and Tile}

\author{\IEEEauthorblockN{HyunSeok Daniel Jang\IEEEauthorrefmark{1},
 Hazem Ibrahim\IEEEauthorrefmark{1}, Rohail Asim\IEEEauthorrefmark{1}, Matteo Varvello,\IEEEauthorrefmark{2}
Yasir Zaki\IEEEauthorrefmark{1}}
\\
\vspace{-0.12in}
\IEEEauthorblockN{
\IEEEauthorrefmark{1}New York University Abu Dhabi,
\IEEEauthorrefmark{2}Nokia Bell Labs}
\\\IEEEauthorblockN{
Email: \{\IEEEauthorrefmark{1}daniel.jang, \IEEEauthorrefmark{1}hazem.ibrahim, \IEEEauthorrefmark{1}rohail.asim, \IEEEauthorrefmark{1}yasir.zaki\}@nyu.edu, \IEEEauthorrefmark{2}matteo.varvello@nokia.com}
}

\maketitle
\begin{abstract}
Bluetooth Low Energy (BLE) location trackers, or ``tags'', are popular consumer devices for monitoring personal items. 
These tags rely on their respective network of \textit{companion devices} that are capable of detecting their BLE signals and relay location information back to the owner.   
While manufacturers claim that such crowd-sourced approach yields accurate location tracking, the tags' real-world performance characteristics remain insufficiently understood. To this end, this study presents a comprehensive analysis of three major players in the market: Apple’s AirTag, Samsung’s SmartTag, and Tile.  Our methodology combines \textit{controlled} experiments---with a known large distribution of location-reporting devices---as well as \textit{in-the-wild} experiments---with no control on the number and kind of reporting devices encountered, thus emulating real-life use-cases. Leveraging data collection techniques improved from prior research, we recruit 22 volunteers traveling across 29 countries, examining the tags' performance under various environments and conditions. %
Our findings highlight crucial updates in device behavior since previous studies, with AirTag showing marked improvements in location report frequency. %
Companion device density emerged as the primary determinant of tag performance, overshadowing technological differences between products. Additionally, we find that post-COVID-19 mobility trends could have contributed to enhanced performance for AirTag and SmartTag. Tile, despite its cross-platform compatibility, exhibited notably lower accuracy, particularly in Asia and Africa, due to limited global adoption. Statistical modeling of spatial errors---measured as the distance betwen reported and actual tag locations---shows log-normal distributions across all tags, highlighting the need for improved location estimation methods to reduce occasional significant inaccuracies.

\end{abstract}


\section{Introduction}
\IEEEPARstart{L}{ocation} tracking technologies have become increasingly prevalent in recent years, with products like Apple's AirTag~\cite{apple_airtag2024}, Samsung's SmartTag~\cite{samsung_smarttag2024}, and Tile~\cite{tile} offering consumers the ability to monitor the location of personal belongings. These devices continuously broadcast Bluetooth Low Energy (BLE) packets and rely on a network of compatible devices to relay their position, raising questions about their accuracy, effectiveness, and potential for misuse (e.g., stalking). While previous research has explored the performance of some of these tags, our study aims to provide a more comprehensive and up-to-date analysis of the location tracking ecosystem.

This paper specifically builds upon the work of~\cite{tags_v1}, which compared the performance of AirTag and SmartTag in both \textit{controlled} and \textit{in-the-wild} experiments. 
The original study utilized custom-developed crawlers for each tag's companion apps (FindMy and SmartThings) to collect detailed location histories reported by devices. Controlled experiments provided insights into tag behavior, e.g., how frequently their location is reported. The tags were deployed in a secluded area alongside Apple and Samsung devices at increasing distances and in a busy campus cafeteria where WiFi connectivity provided an estimate of the surrounding device population.
Complementary in-the-wild experiments explored opportunistic location reporting under varying conditions, e.g., user mobility, population densities, times of day, and days of the week. These experiments involved four volunteers traveling across six countries, each carrying both tags attached to a smartphone equipped with a custom app that logged contextual data such as GPS location and connectivity, etc.

This study addresses key limitations of the original work by (1) expanding the scope of tag devices studied, (2) improving the data collection methodology, and (3) enhancing the representativeness of in-the-wild experiments. The original study was limited to AirTag and SmartTag, which lack cross-platform compatibility and rely solely on their proprietary ecosystems—iOS for AirTag and Samsung devices for SmartTag. To address this, we include Tile, a product compatible with both iOS and Android through its integration with the Life360\footnote{Life360 is a family safety platform which acquired Tile in November 2021. This move extended Tile's network to include Life360 users who opt into location services.}  app~\cite{life360} and the Tile app. This selection allows us to compare \textit{device-dependent} tags (AirTag and SmartTag) with \textit{app-dependent} alternatives (Tile), offering insights into how different approaches to building a location-reporting ecosystem affect performance.

We also address the original study’s reliance on Optical Character Recognition (OCR)~\cite{OCR} for extracting tag locations from companion applications, which introduced potential timestamp errors of up to one minute. In our work, we implement improved crawlers capable of directly parsing cache files for precise, minute-by-minute location monitoring. 
Additionally, we overcome the limited geographical scope and representativeness of the original study, which was conducted in six countries during the COVID-19 pandemic, when social distancing measures likely reduced the density of nearby devices. Our in-the-wild experiments involve 22 volunteers traveling across 29 countries, providing a more realistic and global perspective of tags' performance. By comparing our findings with those of the original study, we evaluate how real-world conditions, including increased device density post-pandemic, influence tag effectiveness. Finally, we leverage the tags' location report dataset to statistically model their distance from ground-truth locations, offering an in-depth evaluation of the methods used by manufacturers to estimate a tag's location. Our key findings are the following:

\vspace{0.05in}
{\noindent}\textbf{Update in Device Behavior.} Controlled experiments revealed significant changes in AirTag behavior, likely due to firmware updates. While the previous study had found AirTags broadcasting weaker BLE signals compared to SmartTags, our new findings indicate that they now transmit at comparable signal strengths to both SmartTags and Tiles, leading to a marked increase in update rate (the number of unique location reports per hour). In the cafeteria experiment, AirTag's update rate scaled with Apple device density, reaching over 45 average updates at peak hours ($>$300 devices)—contrary to~\cite{tags_v1}'s finding of a constant 10-15 updates regardless of device count.

\vspace{0.05in}
{\noindent}\textbf{Importance of Companion Device Density.}
Contrary to earlier findings in~\cite{tags_v1}, our experiment in the wild reveals that the efficacy of location tags is primarily determined by the prevalence of compatible location-reporting devices rather than the specific technology employed. We observe a general correlation between tag performance and the estimated likelihood of encountering companion devices, derived from population density and the country-level mobile market-share. 

\vspace{0.05in}
{\noindent}\textbf{Poor Tile Performance.}
Tile's accuracy is notably lower than that of AirTag and SmartTag, despite its compatibility with both iOS and Android Performance varies by region, with the highest accuracy observed in the Americas and Europe, while accuracy is significantly lower in Asia and Africa. This disparity appears to result from infrequent location updates, likely due to the limited global adoption of Life360/Tile.

\vspace{0.05in}
{\noindent}\textbf{Impact of COVID-19.} 
Following the relaxation of COVID-19 social-distancing measures (2020–2022), public mobility and activity levels gradually returned to pre-pandemic norms, increasing the number of active companion devices available in the environment. This rise in mobility likely contributed to improved location-reporting performance. Compared to the findings in~\cite{tags_v1}, AirTags and SmartTags demonstrate enhanced accuracy, reflecting the greater availability of nearby companion devices as public activity resumed.

\vspace{0.05in}
{\noindent}\textbf{Statistical Modeling for Enhanced Tracking.} 
We apply statistical models to analyze spatial (positional) errors relative to the GPS ground truth. Spatial errors for all tags follow a log-normal distribution, with heavy tails indicating occasional significant deviations. We also find that SmartTag significantly underestimates its margin of error compared to its competitors, raising concerns about the reliability of its location updates in accurately representing true positional uncertainty.

\section{Background and Related Work}
Location tags such as AirTag, SmartTag, and Tile use the BLE protocol~\cite{bluetooth_qprd} to transmit unique identifiers with a range of up to 100 meters. Additionally, SmartTag+ and AirTag support Ultra Wideband
which further extends the range while allowing more precise device localization. Ultra Wideband is only supported by iPhone 11 or later for AirTags, and Samsung Galaxy S21 or later for SmartTag. 

\begin{table}[]
\centering
\resizebox{\columnwidth}{!}{%
\begin{tabular}{|c|c|c|c|}
\hline
\textbf{Tags} & \textbf{\begin{tabular}[c]{@{}c@{}}Companion \\ Devices\end{tabular}} & \textbf{\begin{tabular}[c]{@{}c@{}}Companion \\ App\end{tabular}} & \textbf{\begin{tabular}[c]{@{}c@{}}Opt-in \\ Required\end{tabular}} \\ \hline
AirTag & iPhone \& iPad & \begin{tabular}[c]{@{}c@{}}FindMy\\ (iOS 13+, iPadOS 13+)\end{tabular} & FALSE \\ \hline
SmartTag & \begin{tabular}[c]{@{}c@{}}Galaxy Smartphone \\ \& Tablet\end{tabular} & \begin{tabular}[c]{@{}c@{}}SmartThings\\ (Samsung Android 8+)\end{tabular} & TRUE \\ \hline
Tile & \begin{tabular}[c]{@{}c@{}}iOS/Android \\ Smartphone\end{tabular} & \begin{tabular}[c]{@{}c@{}}Life360, Tile\\ (iOS 16.4+, \\ Android 9+)\end{tabular} & TRUE \\ \hline
\end{tabular}%
}
\vspace{0.1in}
\caption{Location tracking requirements for different BLE tags, including devices capable of detecting BLE signals and relaying GPS data (Companion Devices), required applications and OS versions (Companion App), and whether user opt-in is needed for location reporting.}
\vspace{-0.1in}
\label{tab:tag_companions}
\end{table}

Location tags across different manufacturers generally follow a similar procedure for remote tracking. Each tag must first be registered and associated with an ``owner device'' before use. When a tag moves beyond the proximity of its owner, it enters a ``lost'' state and begins broadcasting BLE signals. These signals are detected by nearby location-reporting nodes, or \textit{companion devices}, which relay their own GPS coordinates to the tag manufacturer. The specific companion devices vary by tag type: for example, AirTags rely on iPhones and iPads, SmartTags use Samsung mobile devices, and Tiles are compatible with iOS and Android smartphones running the Life360 or Tile applications. Using the GPS data from these companion devices, the manufacturers estimate the lost tag’s location and report this information back to the owner via \textit{companion applications}. 
Table~\ref{tab:tag_companions} summarizes the companion devices and applications for AirTag, SmartTag, and Tile. Note that for SmartTag and Tile, the users are required to ``opt-in'' within their respective companion apps to activate location tracking services.           

Despite privacy safeguards, measures to deter malicious tracking had been insufficient, as discussed in~\cite{heinrich2021devicessecurityprivacyapples}. Previously, vendors only alerted users if an unpaired tag from the same manufacturer was detected nearby for an extended period, leaving them vulnerable to cross-vendor misuse (i.e., AirTags used to stalk Samsung users). 
Earlier solutions included Apple's ``Tracker Detect'' \cite{apple_tracker_detect}, an Android app for manually scanning for AirTags, and Heinrich et al.'s proposed system that automatically flags the same AirTag encountered in three separate locations within 24 hours~\cite{heinrich_airguard}. Briggs et al.~\cite{briggs_ble_doubt} extended this design to support generic tags, not just AirTags. However, these methods were limited by MAC address randomization~\cite{apple_wifi_privacy}, which causes tags to appear as new devices to third-party apps over time. 
To overcome these challenges, Apple and Google introduced a joint specification for ``unwanted tracking alerts''~\cite{ietf_unwanted_tracking_2024} in May 2024. This allows mobile devices across both iOS (17.5 and later) and Android (6.0 and later) to notify users of any unpaired tags repeatedly broadcasting BLE signals over an extended period, indicating potential tracking behavior~\cite{apple_unwanted_tracking}. The list of BLE trackers that support this specification  currently includes AirTag, SmartTag, Tile, Chipolo, Eufy, and Pebblebee. 

Much of the prior research has focused on identifying security vulnerabilities in the firmware and protocol design of BLE-based location tags. Martin et al.~\cite{Martin_2019} reverse-engineered Apple's BLE Continuity protocol, uncovering unencrypted transmissions that exposed device activity and user-identifying information. Roth et al.~\cite{Roth_2022} highlighted weaknesses in AirTags' hardware protections, using voltage glitching to clone devices and demonstrate over-the-air manipulation capabilities. Similarly, Yu et al.~\cite{yu2022privacyanalysissamsungscrowdsourced} scrutinized the Offline Finding (OF) protocol of Samsung SmartTags, revealing vulnerabilities in BLE payload generation, privacy ID randomization, and cryptographic implementations. Givehchian et al.~\cite{Givehchian_2022} explored the privacy implications of BLE protocol devices, such as location tags, demonstrating that physical-layer identification is possible but often unreliable. 

In contrast, research on the global performance of location tags in real-world scenarios remains limited.
Hernández et al.~\cite{hernandez_2023} studied the efficiency of finding AirTags and Tile tags on a university campus through real and simulated experiments, where they model the probability of locating a tag based on the flow rate of individuals carrying compatible smartphones within the tag's detection range. They showed that AirTags had a range of 10–30 meters (consistent with~\cite{tags_v1}) and that in populated areas, both tags relayed their location within one hour 98\% of the time.

Ibrahim et al.~\cite{tags_v1} examined the performance of two popular Bluetooth Low Energy (BLE) location tags—Apple's AirTag and Samsung's SmartTag—through controlled experiments and real-world scenarios. The study compared the tags' accuracy and responsiveness in reporting locations, considering factors such as population density, user mobility, and the number of nearby reporting devices. They found that both tags perform similarly, typically locating within 100 meters in about 10 minutes. Despite Samsung's more aggressive update strategy, the paper claimed that real-time use of either tag for stalking is impractical, though half of a victim's movements could be backtracked with a one-hour delay. This paper serves as an extension to~\cite{tags_v1} by addressing the following limitations: 

\begin{enumerate}[left=0pt]
    \item The scope of~\cite{tags_v1} was limited to AirTag and SmartTag,  both of which lack cross-platform compatibility, limiting insights into tags that work across multiple operating systems. \textit{This extension includes Tile due to its longstanding presence since 2012 and its recent expansion through integration with Life360's user base in 2021.}
    \item The crawling methodology relied on OCR to extract tag locations from their respective companion applications. This approach introduced potential errors of up to one minute in the reported location timestamps. \textit{We enhance this methodology to achieve second-level granularity in location updates without using OCR.}
    \item The ``in-the-wild'' data collection was conducted in six countries during the COVID-19 pandemic (March to August 2022), when social distancing measures likely reduced the density of nearby companion devices, impacting the representativeness of the results. \textit{We conduct a more expansive real-world measurement campaign across 29 countries after such regulations were lifted.}
\end{enumerate}

\section{Methodology}
\label{sec:method}

\begin{figure}[t]
    \includegraphics[width=\linewidth]{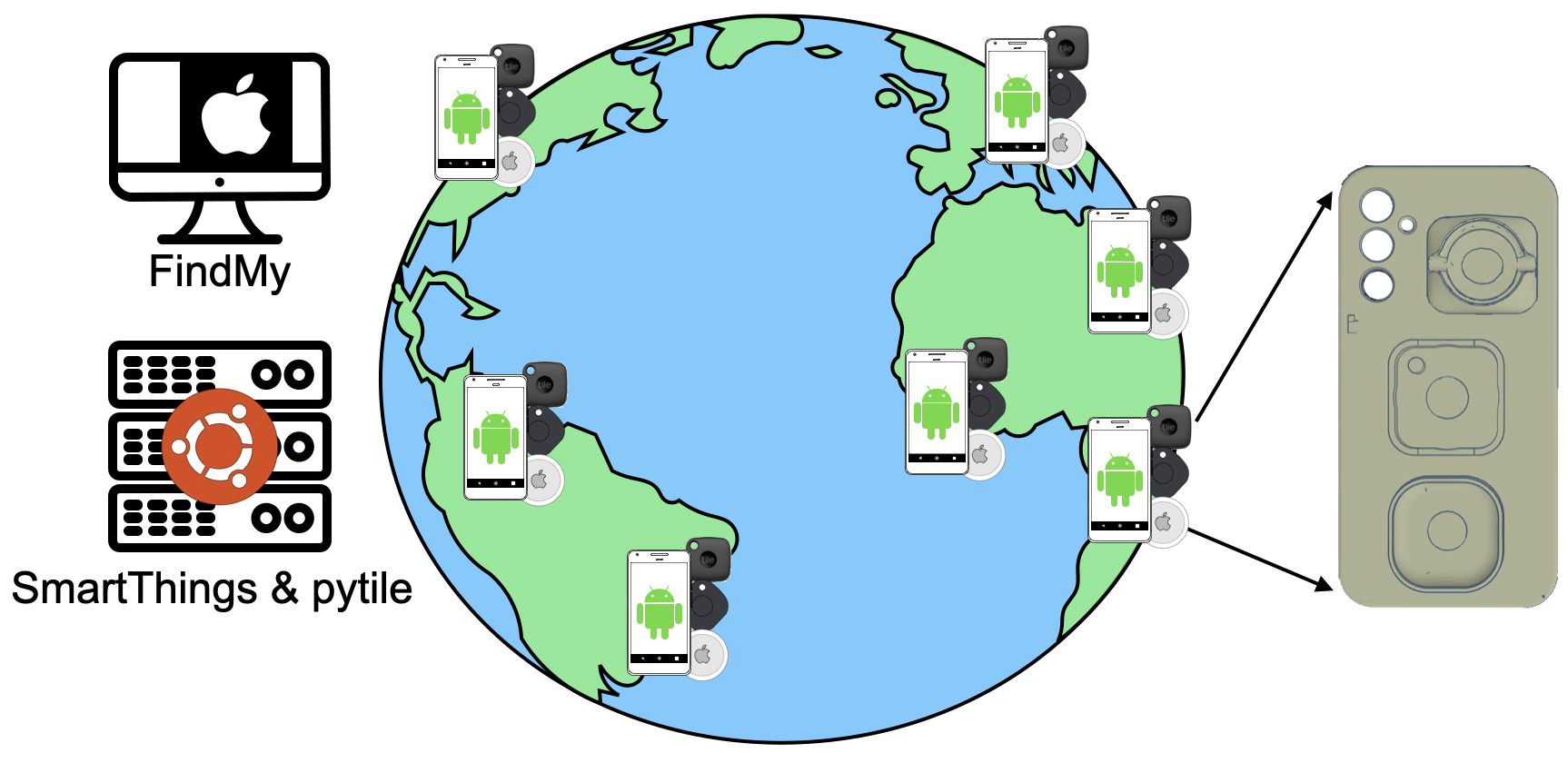}
    \caption{On the left, two data collection servers (MacOS and Ubuntu) run the FindMy, SmartThings, and pytile crawlers. On the right, a sample 3D mockup of our vantage point, a Galaxy A34 equipped with three tags.} 
    \label{fig:platform_visual}    
\end{figure}

\begin{figure}
    \centering
    \includegraphics[width=0.85\linewidth]{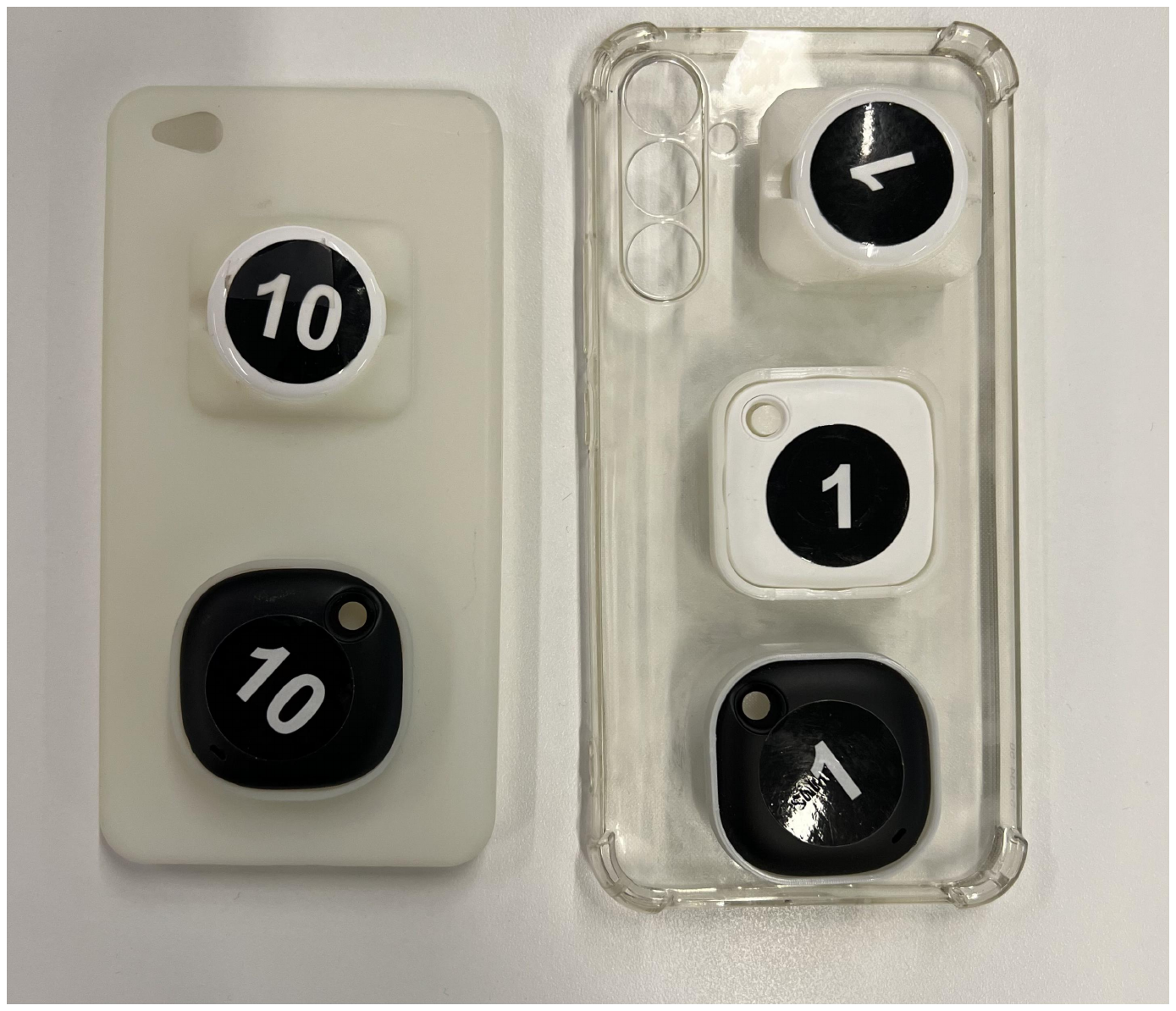}
    \caption{3D-printed cases for the Redmi Go (left) and Samsung Galaxy A34 (right), used as vantage points.} 
    \label{fig:vantage_point_cases}    
\end{figure}

Our study focuses on three prominent BLE trackers currently available in the market: Apple's AirTag, Samsung's SmartTag, and Tile (Mate). Figure~\ref{fig:platform_visual}  provides a visual overview of our platform for analyzing the tags of interest. Our approach centers on two key components: data collection servers and mobile vantage points. The former runs on MacOS and Linux, hosting custom crawlers that continuously query and record location data from the tags' companion apps (FindMy, SmartThings, and Tile). The mobile vantage points are Android smartphones equipped with a custom case housing all three tag types, which we deployed to volunteers across numerous countries. 

\subsection{Tag Crawlers}
\label{sec:method:crawlers}
The real-time location data of different tags are provided in each of their companion app: FindMy for AirTag (iOS and MacOS exclusive), SmartThings for SmartTag (Samsung Android exclusive), and Tile/Life360 app (available on both iOS and Android) for Tile. Since there are no official APIs published by the manufacturers, we designed custom ``crawlers'' that automate location monitoring for each type of the tags deployed during the data collection stage. Previous crawlers for AirTags and SmartTags, as used in~\cite{tags_v1}, relied on automated interactions to ``pin'' their locations to Apple Maps and Google Maps, respectively. Next, Optical Character Recognition (OCR) was used to read coordinates and timestamps from the maps, calculating the reporting time based on the crawling epoch and the ``last seen'' time (e.g., ``X minutes ago'') displayed in their companion apps. This approach introduced potential errors of up to one minute for 47\% of location reports. To address this limitation, we developed a more reliable and precise method for collecting location data, capturing both coordinates and exact reporting timestamps with second-level granularity.

\subsubsection{AirTag Crawler}
\label{sec:method:crawlers:airtag}
Our AirTag crawler leverages an open-source bash script~\cite{icepick3000_2024} designed to run on macOS with an authenticated Apple account. During each execution, the script opens the FindMy Application and parses its cache file\footnote{\texttt{/Library/Caches/com.apple.findmy.fmipcore/Items.data}}, which stores the most recent location information of devices owned by the Apple account. 
During our experiments, the maximum number of items tracked by FindMy was limited to 16~\cite{findMyMax}. Hence, to continuously monitor 22 AirTags deployed simultaneously, we distributed the workload across two macOS servers, each executing the crawling script at one-minute intervals. 

\subsubsection{SmartTag Crawler}
\label{sec:method:crawlers:smarttag}
We reverse-engineered SmartThings to investigate how it stores the location data of SmartTags registered by the authenticated Samsung account, given the lack of relevant public documentation. Upon decompiling the app's Android Package (apk) into its source code and resources, we identified a cache file\footnote{\texttt{com.samsung.android.oneconnect/shared\_prefs/FME\_SEL\\ECTED\_DEVICE.xml}} that contains detailed metadata---including last reported location---for the device last displayed on screen by SmartThings. Building on this finding, we developed a custom automation script that leverages the Android Debugging Bridge (ADB) to iterate through each registered SmartTag. At each iteration, the script prompts the app to refresh the cache file for the selected tag, then extracts the precise location coordinates and timestamps directly from the app's internal data structure. To ensure minute-level location probing for each SmartTag, we simultaneously run the SmartTag crawlers across 4 rooted Samsung devices (A34 5G) that were connected to a  Linux server via USB. 

\subsubsection{Tile Crawler}
\label{sec:method:crawlers:tile}
We developed a Python script leveraging the pytile library~\cite{bachya_pytile}, which interfaces with a custom API for Tile.
The script performs two main functions: it authenticates with the Tile service and the retrieves location information of all Tile tags linked to the account. In our setup, we used a single Tile account to manage all 22 devices used in the study. We automated the script to run every minute on a Linux server, but the implementation can run on any operating systems since it has no platform-specific dependencies.

\subsection{Vantage Point}
\label{sec:method:vantage_point}
Our vantage points consist of rooted Android smartphones. Ibrahim et al.~\cite{tags_v1} employed Xiaomi Redmi Go devices to compare the performance between AirTags and SmartTags. However, the Redmi Go's limited 3,000~mAh battery frequently led to rapid depletion. To address this issue, we introduced the Samsung Galaxy A34 in this extension, which offers a larger 5,000~mAh battery. Additionally, the Galaxy A34's larger dimensions allowed us to incorporate the additional Tile tag. Each tag was securely mounted on a custom-designed, 3D-printed cover attached to the back of the smartphone. Figure~\ref{fig:vantage_point_cases} shows the cover for Redmi Go used in~\cite{tags_v1} on the left, followed by the cover for the Galaxy A34. The devices were further configured to avoid reporting the location of any of the attached tags; for Samsung devices, we disabled ``Offline finding'' option in the Find My Mobile settings to exclude the device from SmartThings network. Additionally, we verified that neither the Life360 nor Tile applications were installed. Note that Android devices cannot be involved in Apple's FindMy network, thus precluding any inadvertent location reporting of AirTags.   

To precisely document the ground truth location of vantage points over time, we extend the custom Android application used in~\cite{tags_v1}. Manually installed in our devices, this app utilizes Android's LocationRequest API~\cite{android_location_request} to record GPS location at one-second intervals. For this study, we configured the API at \texttt{QUALITY\_HIGH\_ACCURACY} mode, which prioritizes location accuracy at the expense of increased power consumption—for example, by using Wi-Fi-based locations instead of cell-based locations. 
This approach provided finer-grained ground truth location data compared to the previous study, which employed \texttt{QUALITY\_BALANCED\_POWER\_ACCURACY} mode and logged locations at 5-second intervals to accommodate the battery limitations of the Redmi Go. The app employs a buffering system which stores the location data for up to five minutes before transmitting it to our server via a POST request when internet connection is available. If connectivity is unavailable, the data remains buffered until a connection can be established, ensuring no loss of information. We compare this ground truth against the locations reported by our crawlers to assess tags' performance.

\subsection{Dataset Description}
\label{sec:method:description}

\begin{figure}[t]
\centering
\includegraphics[width=\linewidth]{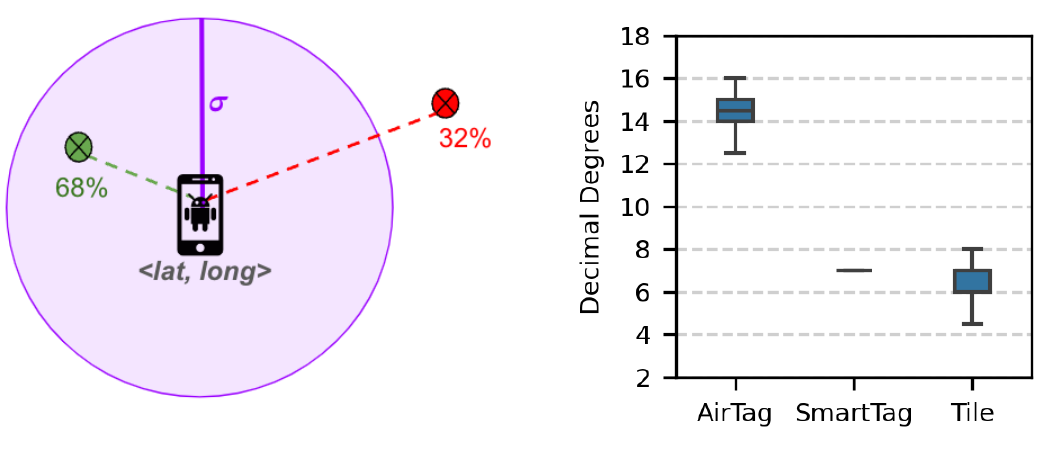}
\caption{On the left, an Android GPS location reported at a specific timestamp, showing latitude, longitude, and $\sigma$ at 68\% confidence interval. On the right, precision of reported location coordinates, measured in decimal degrees, collected by our crawlers for different tag types.}
\label{fig:data_description}
\end{figure}
Each location update from GPS or BLE tags includes four data points: $<$timestamp, latitude, longitude, $\sigma$$>$. As shown in Figure~\ref{fig:data_description} (on the left), $\sigma$ represents the radius of a circle centered around the reported coordinates and is guaranteed to contain the true position within a certain probability. In essence, $\sigma$ serves as an estimate of the location report's margin of error.
Android documentation specifies that GPS reports provide $\sigma$ at the 68th percentile, meaning there is a 68\% probability that the actual location lies within this radius. BLE tag manufacturers have not disclosed the confidence interval of $\sigma$ for their devices. Note that vendors %
label $\sigma$ differently
(e.g., ``horizontal accuracy'' used by Android and Apple, ``horizontal uncertainty'' used by Samsung, ``accuracy'' used by Tile); however, we use $\sigma$ throughout this paper for clearer referencing.

\textit{Coordinate precision}—measured by the number of decimal places in location coordinates—is an important indicator of location data quality. As shown in Figure~\ref{fig:data_description} (on the right), location reports from our AirTags crawler demonstrated the highest precision, with over 88\% of coordinates having 14 or more decimal degrees. 
This consistently high precision suggests that Apple's companion devices provide highly detailed location data. In comparison, both SmartTag and Tile reports averaged seven decimal degrees, offering a margin of error of approximately 1.11 cm at 0.0000001 degrees—still remarkably precise. 
However, in very rare cases (approximately 0.75\%), Tile location reports had fewer than five decimal degrees, potentially introducing errors of up to 11.1 meters. While such occurrences were extremely infrequent, they suggest occasional precision limitations. Overall, the location reports collected from our tag crawlers are sufficiently precise to support our analysis of their location tracking capabilities.

\section{Data Collection}
\label{sec:data_collection}
Tags' efficacy for locating tracking depends on two factors: 1) the technology adopted, and 2) the probability of encountering a companion device, \textit{i.e.,} a Samsung or Apple smartphone with enabled Bluetooth, GPS location, and data connectivity. 
While the reach of the technology can be studied in a lab, the opportunistic encountering of a reporting device requires experiments in-the-wild to account for realistic conditions. 
To this end, we adopt the experiments in~\cite{tags_v1} with our enhanced crawlers. 
First, we measure the \textit{Received Signal Strength Indicator (RSSI)} in an isolated area to evaluate the quality of communication between tags and companion devices (Section~\ref{sec:data_collection:rssi}). 
Next, we deploy tags in a busy campus cafeteria to compare their location update strategies, as indicated by their \textit{update rate} (Section~\ref{sec:data_collection:update_freq}). Finally, we conduct an expanded in-the-wild experiment with a larger number of volunteers and significantly broader geographic coverage compared to the original study (Section~\ref{sec:data_collection:in_the_wild}).

\subsection{Isolated Area}
\label{sec:data_collection:rssi}
We replicated the experiment in~\cite{tags_v1} in a secluded area, 300 meters away from any building, to isolate the performance of our tags from external interference. This controlled environment contained only our tags and test phones. We positioned three smartphones at varying distances—directly adjacent to, 10 meters, and 50 meters away from each tag type. These devices captured and measured the RSSI of BLE beacons emitted by the tags. SmartTags include the device local name (``Smart Tag'') within their BLE advertisements. For AirTags, we leveraged the fact that accessories in the FindMy network share the first 6 bytes (``1EFF4C001219'') when advertising BLE packets in \textit{separated state} (i.e., when distant from the owner device)~\cite{catley2024airtag}. Tile beacons were identified by filtering BLE packets for Tile's 16-bit Service UUID as defined in Bluetooth specifications~\cite{bluetooth2024assigned,ariccio_2021}.

\subsection{Campus Cafeteria}
\label{sec:data_collection:update_freq}
We placed three of each tag types for a week in a university cafeteria which serves approximately 1,200 people daily from 7am to 11pm. Throughout the experiment, we ran our custom crawlers while collaborating with the university's IT department to monitor the presence of tag-compatible devices nearby the deployed tags. Specifically, we leveraged the department's shared data which identify the manufacturer of mobile endpoints connected to the cafeteria WiFi. This allowed us to obtain an exact count of AirTag-compatible devices (iPhones, iPads) present in the area. Since the FindMy network is activated by default on Apple devices, this count directly represented potential AirTag reporters. 

While we could also obtain the total number of Samsung devices, this would overestimate SmartTag-compatible devices since users must opt-in to \textit{Offline finding}. To refine our count, we filtered network traffic with destination containing \texttt{chaser-eu02-euwest1.samsungiotcloud.com}, \\ which is the domain URL for geolocation reports of SmartTags generated by non-owner devices (with ``eu02-euwest1'' representing the region subdomain)~\cite{yu2022privacyanalysissamsungscrowdsourced}. By counting unique devices sending requests to this domain, we achieved a more accurate tally of SmartTag-compatible devices. The percentage of Samsung devices which opted-in for SmartThings ranged from 11\% to 26\% per hour during our experiment. 

For Tile, we counted the unique number of devices sending API requests to URLs containing \texttt{life360.com} or \texttt{tile-api.com}, as Tile locations can be reported by devices with either the Life360 or Tile app installed (and where users have opted-in to report locations of Tile tags).

The collected data was aggregated into time-segmented counts of each tag-compatible device type, ensuring complete anonymization of individual users. This approach allowed us to estimate the number of active location-reporting devices for each tag type present in the cafeteria at different times. We acknowledge that devices not connected to WiFi were not captured in our data, although the cafeteria's poor cellular coverage likely minimized this issue. 

\subsection{In-the-wild}
\label{sec:data_collection:in_the_wild}

\begin{table*}
\begin{center}
\begin{tabular}{| c | c | c | c | c | c | c | c | c | c | c }
\hline
\textbf{Campaign} & \textbf{Device} & \textbf{Duration} & \textbf{Distance (km)} & \textbf{Countries} & \textbf{Cities} & \textbf{\# AirTag} & \textbf{\# SmartTag} & \textbf{\# Tile} \\
\hline
Original & Redmi Go & Mar-Aug 2022 & 9,378 & 6 & 20 & 21,081 & 3,595 & N/A \\ 
\hline 
Extended & Galaxy A34 5G & Dec 2023 - Jan 2024 & 177,621 & 29 & 123 & 70,974 & 10,5426 & 8,440 \\ 
\hline 
\end{tabular}

\end{center}
\caption{Summary of two measurement campaigns. \# refers to the number of unique location updates for each tag.}
\label{tab:campaigns}
\end{table*}

We perform in-the-wild measurements to assess the tags performance in real-world scenarios. 
In~\cite{tags_v1}, Redmi Go devices equipped with AirTags and SmartTags were deployed as vantage points by four volunteers between March and August 2022, covering 9,378 kilometers across six countries and 20 cities. To include Tile as an additional tag and address the Redmi Go's limited battery capacity, we conducted a new measurement campaign from December 2023 to January 2024.  
In this campaign, updated vantage points—Samsung Galaxy A34 5G encasing all three tags (see Figure~\ref{fig:vantage_point_cases})—were distributed to 22 volunteers, who collectively traveled 177,621 kilometers across 29 countries and 123 cities. 
For clarity, we will refer to the two measurement campaigns as the \textit{original campaign} and the \textit{extended campaign}, respectively.
A comparison of the two campaigns is summarized in Table~\ref{tab:campaigns}.
For a more detailed breakdown of experiments conducted in each country, refer to Appendix~\ref{appendix:campaign_breakdown}.

\begin{figure}
    \centering    
    \includegraphics[width=1\columnwidth]{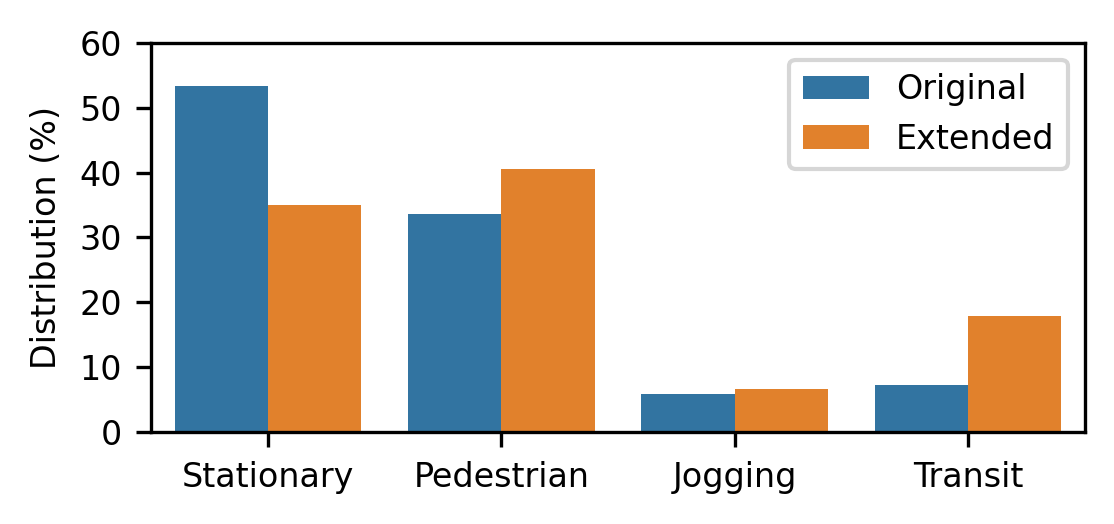} 
    \caption{Distribution of mobility levels between vantage points in the original vs. extended campaign.} 
    \label{fig:mobility_distribution}    
\end{figure}

We further compare the mobility of vantage points based on their GPS dataset, categorizing their speed within every 10-minute window into four levels: stationary ($<$ 0.2 km/h), pedestrian (0.2-6 km/h), jogging (6-12 km/h), and transit ($>$ 12 km/h). 
Figure~\ref{fig:mobility_distribution} plots the distribution of mobility patterns between the two campaigns. 
In the original campaign, more than half of the measurements were taken when volunteers were stationary, while this proportion decreased to 35\% in the extended campaign. Although jogging remained infrequent in both ($<$ 7\%), the extended campaign showed a 7\% increase in pedestrian activity and a 10.6\% increase in transit activity compared to the original.
These changes reflect the influence of the COVID-19 pandemic during the original campaign, where social-distancing policies limited participants' mobility. The relaxation of these policies during the extended campaign facilitated greater outdoor activity, potentially improving the ``in-the-wild'' tag performance %
compared to the original study. 

Throughout both measurement campaigns, participants were instructed to carry the vantage points with them as much as possible during their daily activities and travels, limiting interactions to necessary actions such as charging. Additionally, we implemented a filtering process that excluded data recorded within a 300-meter radius of each participant's temporary ``home'' locations, which include not only permanent residences but also hotels, hostels, or any other accommodation where participants stayed overnight. This was to prevent potential data skew that could occur if a nearby device (such as a neighbor's or family member's phone) consistently reported a tag's location. After applying this filter, we retained 24.7\% of the total data collected. We also exclude data from South Korea in our analysis of AirTag, as FindMy support is not expected in the region until Spring  2025~\cite{engadget_findmy2024}. 

We use the dataset collected to analyze tags in two complementary aspects: a high-level examination of accuracy aggregated across relevant social-geographical factors (Section~\ref{sec:accuracy}), and a more granular investigation, quantifying the expected reliability of individual location updates (Section~\ref{sec:reliability}).

\section{Controlled Benchmarks}
\label{sec:control}

\subsection{Bluetooth Signal Strength}
\label{sec:exp:rssi}
\begin{figure}
    \centering    
    \includegraphics[width=0.9\columnwidth]{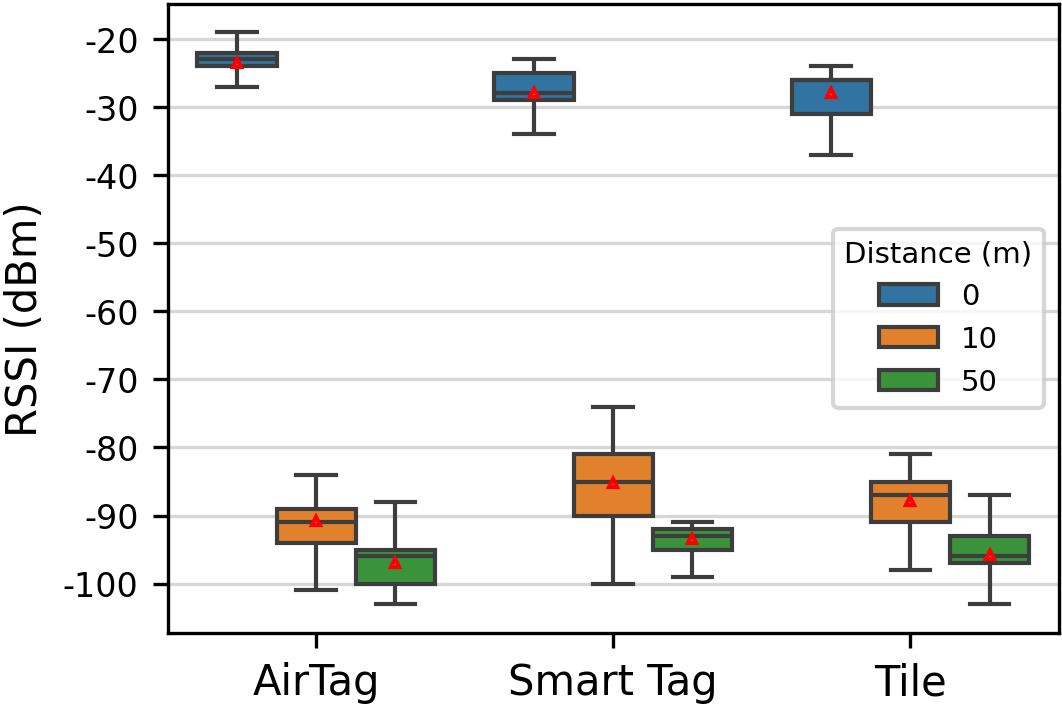} 
    \caption{RSSI strength for each tag type at varying distances. The red triangles mark the statistical average.} 
    \label{fig:rssi}    
\end{figure}

Figure~\ref{fig:rssi} shows that all three tag types exhibited comparable levels of Received Signal Strength Indicator (RSSI). At most distances, the average pairwise differences between the tags were less than 3 dBm, a  variation generally considered insufficient to meaningfully impact Bluetooth connectivity. The notable exception occurred between the AirTag and SmartTag at 10 meters, where the average RSSI difference exceeded 5 dBm. This larger discrepancy suggests that the SmartTag's beacon was received with approximately double the power of the AirTag, potentially influencing device detection range and signal stability at this distance.

Despite the exception at 10 meters, results for the AirTag significantly deviate from those reported in~\cite{tags_v1}. The previous study found that AirTag beacons were consistently received with an RSSI approximately 10 dBm lower than SmartTag beacons at all distances up to 50 meters, where the AirTag signals were no longer detected. We conjecture that this improvement may be attributed to firmware updates released for AirTags since the earlier study. The AirTags used in~\cite{tags_v1} were equipped with firmware version 1.0.301, whereas our tags operated on version 2.0.73. Note that these updates occur automatically when an AirTag is within Bluetooth range of its paired iPhone~\cite{iclarified2024}, making it challenging to control for specific firmware versions during our experiments.   

\begin{figure*}
    \centering    
    \includegraphics[width=\textwidth]{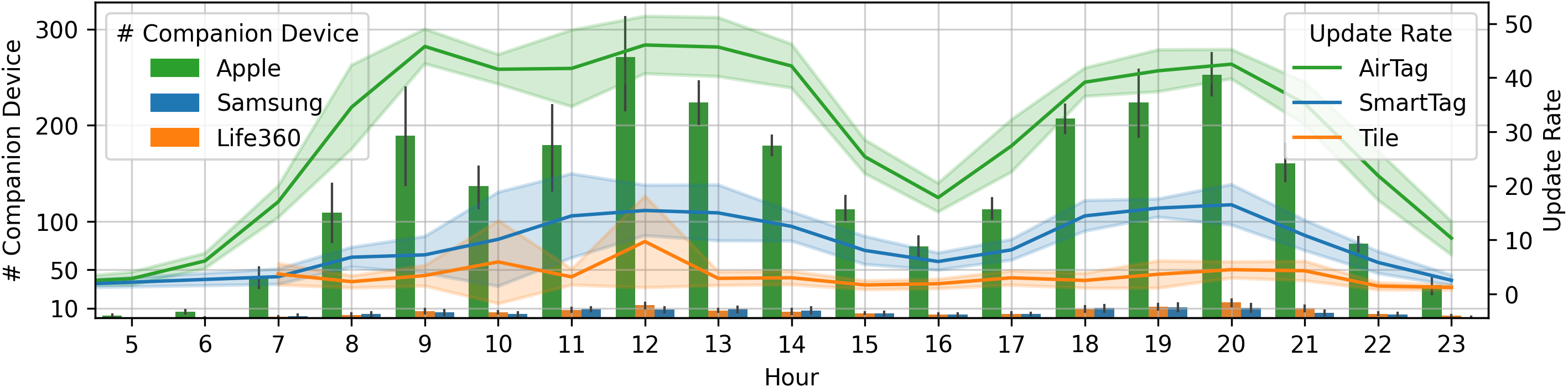} 
    \caption{Update rates of Tags at different times of day in a campus cafeteria.} 
    \label{fig:update_rate_tod}    
\end{figure*}

\subsection{Update Rate}
\label{sec:exp:update_freq}

\begin{figure*}[t]
\centering
\begin{minipage}[t]{1\columnwidth}
    \includegraphics[width=1\columnwidth]{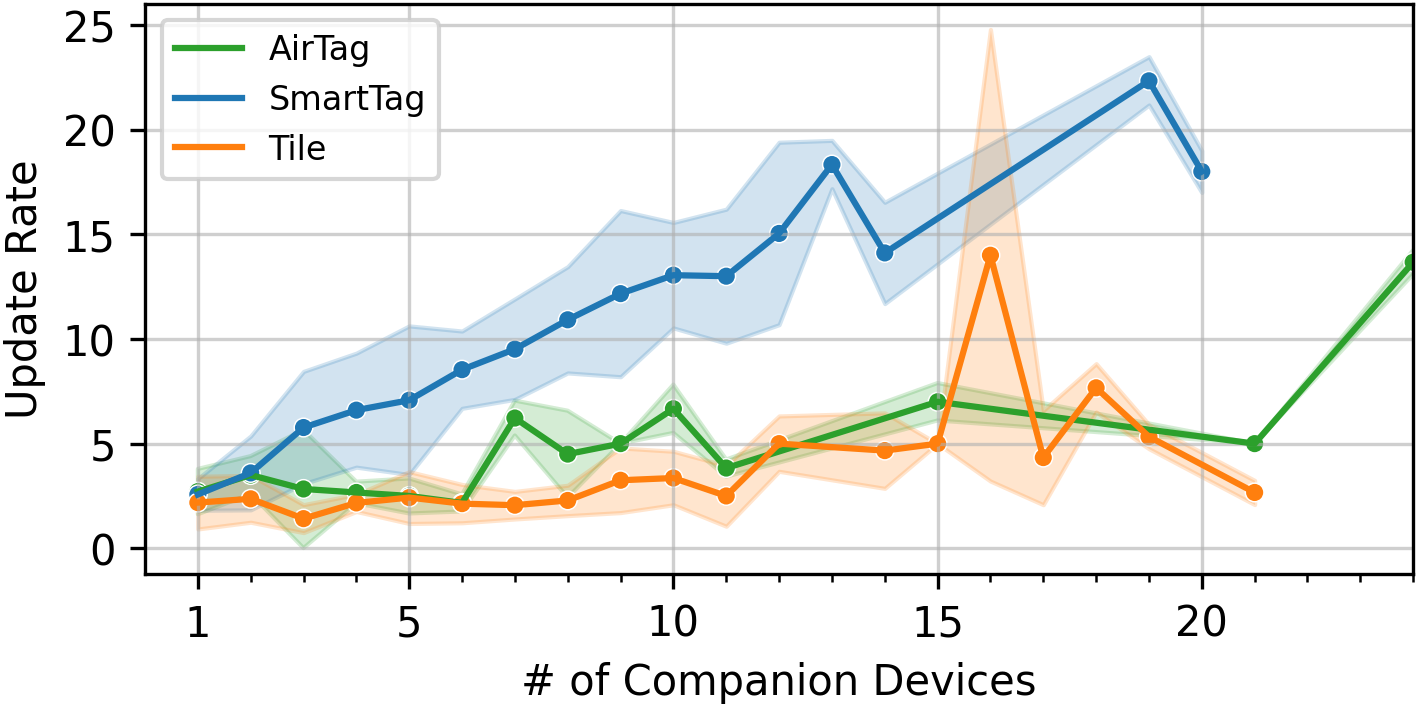}
        \caption{Tag update rates as a function of compatible-devices ($<$25) per hour.} 
    \label{fig:update_rate_below_25}   
\end{minipage}
\hfill
\hfill
\begin{minipage}[t]{1\columnwidth}
    \includegraphics[width=1\columnwidth]{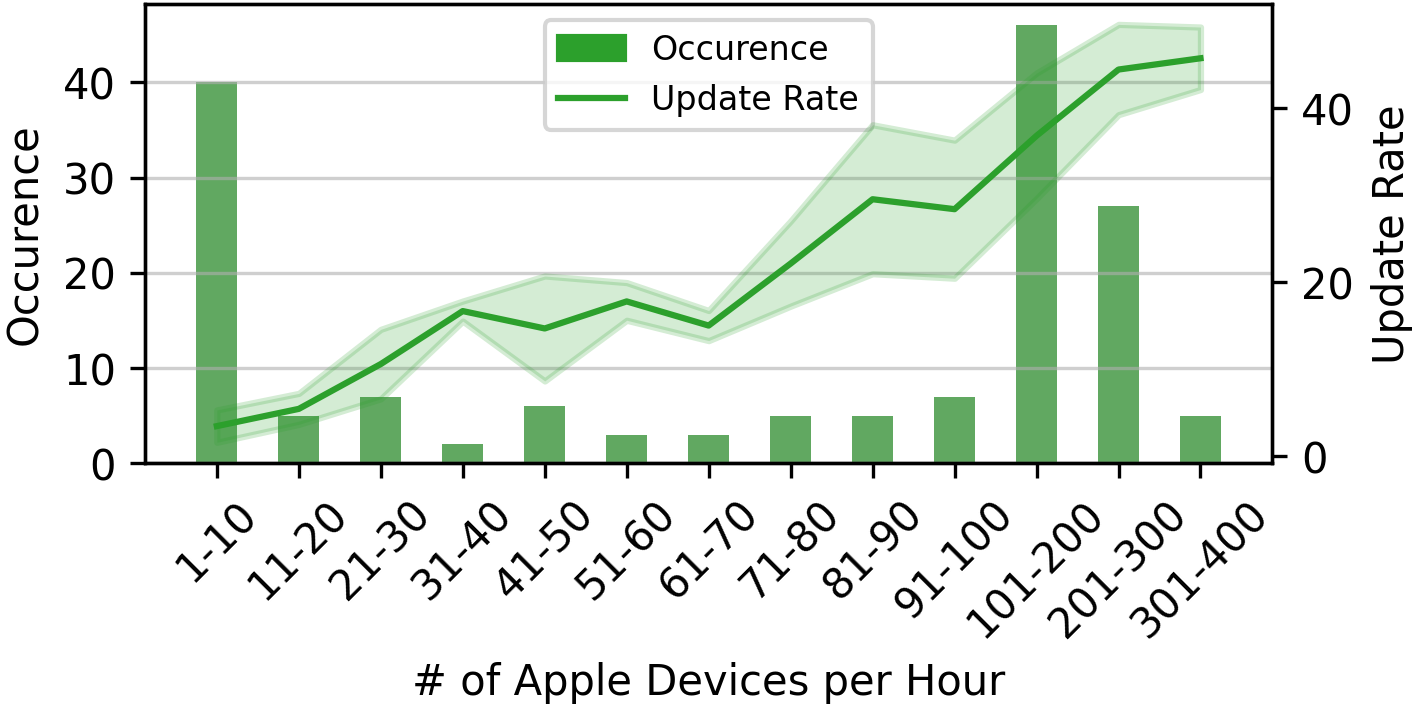}
        \caption{AirTag update rate as a function of the likelihood to have N companion devices (iPhone, iPad) within one hour.} 
    \label{fig:airtag_update_rate}    
\end{minipage}
\end{figure*}
Figure~\ref{fig:update_rate_tod} illustrates the update rate as a function of surrounding compatible devices for AirTag, SmartTag, and Tile. The graph presents, for each hour, the average tag update rate and device count over the 4-day experiment period. Shaded areas and error bars indicate the standard deviation for each metric. The figure shows significant differences in performance and device prevalence among the three tag types. AirTags demonstrate the highest update rate, peaking at around 50 updates per hour during busy lunch periods. This performance correlates with the consistently high number of Apple devices, which often exceed 100 in working hours. In contrast, both SmartTag and Tile had far fewer companion devices present, generally less than 10 in most hours. Despite this limitation, SmartTag maintains 10-20 updates per hour, while Tile typically falls below 10 updates per hour. This data suggests that SmartTag's network may be more aggressive in reporting locations compared to that of Tile.

To further analyze these results, we examine the update rate as a function of the number of companion devices present within an hour. Figure~\ref{fig:update_rate_below_25} compares the update rates for each tag type when there are fewer than 25 companion devices nearby. Note that our controlled experiment yields limited data for SmartTag and Tile at higher companion device counts. In the figure, circular markers represent the average update rate at each measured device count, while shaded regions indicate the standard deviation where data is available.
Our observations reveal that SmartTag employs the most aggressive location reporting strategy. Its update rate generally increases with a higher number of companion devices, peaking at 23 updates per hour with 19 devices present. Notably, SmartTag requires eight companion devices to achieve more than ten location updates per hour. In contrast, AirTag and Tile converge to similar update rates, consistently remaining below ten updates per hour when fewer than 21 companion devices are nearby. An interesting exception occurs for Tile at 16 companion devices, where the update rate often ``spikes'' above 20. %

Figure~\ref{fig:airtag_update_rate} illustrates the relationship between AirTag update rates and larger numbers of companion devices nearby. The bar plots represent the frequency of occurrences for each range of Apple device counts. Our analysis reveals that AirTag's update rate initially plateaus between 10-20 updates per hour when 30-70 Apple devices are present. However, as the number of Apple devices further increases, the update rate continues to rise, averaging over 45 updates per hour with more than 300 companion devices nearby. This finding contrasts with the results reported in~\cite{tags_v1}, which suggested a consistent update rate of 10-15 across all Apple device counts.

\subsection{Discussion}
Our findings show that all three tags emit BLE packets at similar RSSI, making it unlikely that performance differences are attributable to variations in signal strength. Our subsequent cafeteria experiment demonstrates that SmartTag adopts a more aggressive update strategy compared to the other tags when the number of nearby companion devices is limited ($\leq$20 devices). 
As noted in~\cite{tags_v1}, this behavior is likely a response to the requirement for Samsung devices to opt-in to the SmartThings network to report the locations of other SmartTags—a barrier not present for Apple's FindMy. While Tile faces a similar opt-in requirement as SmartTag, its location update strategy remains as passive as Apple's. 

Despite its relatively passive strategy, the AirTag's update rate was more than double that of the SmartTag throughout the controlled measurements in the cafeteria, driven by the significantly higher number of companion devices in proximity. This highlights the fundamental reliance of crowd-sourced location updates on companion device density, which may play a more critical role in tag performance than the update strategies employed.

Finally, compared to~\cite{tags_v1}, we observe a significant improvement in AirTag's update rate when surrounded by a large number ($\geq$70) of companion devices, which may be attributed to recent firmware patches as described in~\ref{sec:exp:rssi}. Additionally, advancements in the method for identifying companion devices could have played a role. Device identification had previously relied on analyzing traffic logs associated with Apple or Samsung application IDs. In this extension study, however, the IT department utilized advanced device profiling features from wireless LAN controllers~\cite{Cisco_2024}, which can identify device types with greater accuracy, including specific models (e.g., iPhone 12). This improvement in device identification may have contributed to the differences observed between our findings and those in the original study.  

\section{Location Tag Accuracy}
\label{sec:accuracy}
This section analyzes tags accuracy, incorporating data collected from the two in-the-wild campaigns outlined in Table~\ref{tab:campaigns}.  Intuitively, a tag's location report can be evaluated along two dimensions: spatial error (distance from the ground truth), and temporal error (responsiveness to location changes) Section~\ref{sec:accuracy:methods} outlines the methodology used to assess tag accuracy based on these dimensions, while Section~\ref{sec:accuracy:radius_accuracy} explores the relationship between them in greater detail.

In Section~\ref{sec:accuracy:covid}, we compare tag accuracy between the two measurement campaigns, shedding light on the potential impact of COVID-19 on the measurement reported in the original study. 
Finally, Section~\ref{sec:accuracy:global} shows findings unique to this extension, presenting the regional characteristics in accuracy observed across tag types, as well as implications for Tile's (poor) performance. 

\subsection{Evaluation Metric}
\label{sec:accuracy:methods}
We follow the methodology in the original paper to calculate the accuracy for each pair of vantage point and tag type, accounting for both the physical proximity (spatial error) and responsiveness (temporal error) of location updates. 
Reported locations are grouped into X-minute intervals for each vantage point. For each interval, we compute the haversine distance~\cite{haversine} between the GPS position and the tag's reported position. A ``hit'' is recorded if the distance falls within a specified radius. Accuracy is then computed as the percentage of hits across all intervals. This provides a direct quantitative insight for real-world tracking: \textit{what is the likelihood that a tag accurately updates its location within X minutes?}
\begin{figure}[!t]
    \centering
    \includegraphics[width=0.8\linewidth]{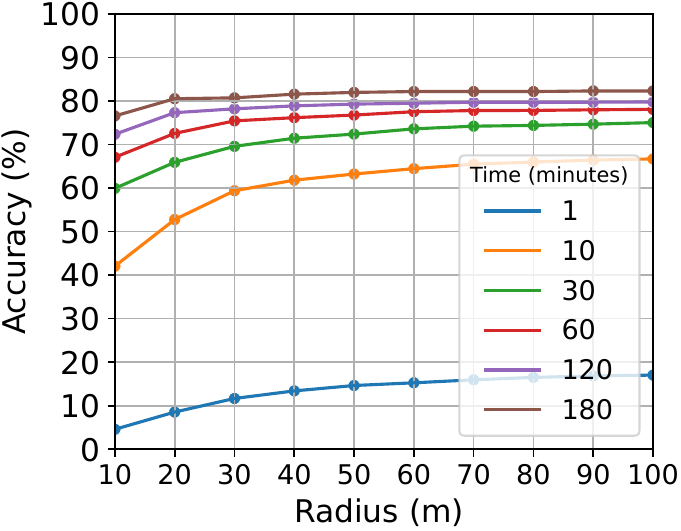} 
    \caption{Combined accuracy of tags vs. radius across different time windows.} 
    \label{fig:radius_sweep}    
\end{figure}

\begin{figure*}
    \includegraphics[width=\textwidth]{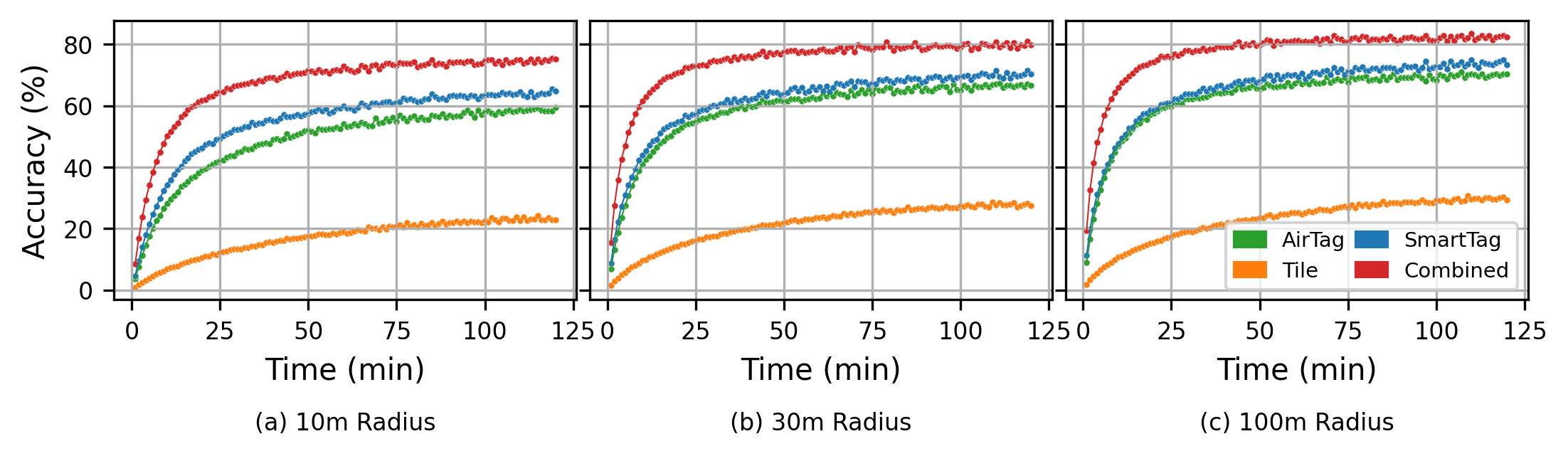} 
    \caption{Tag accuracy when considering different radii as ``hits''. For each tag type, the lines represent the percentage of location reports falling within (a) 10 meters, (b) 30 meters, and (c) 100 meters away of GPS ground truth, plotted against the minutes elapsed since the GPS coordinate was recorded.} 
    \label{fig:accuracy_time_sweeps}    
\end{figure*}

\subsection{Spatial vs. Temporal Error}
\label{sec:accuracy:radius_accuracy}
We begin our analysis by investigating each tag's temporal error within a specific radius from the ground truth (spatial error). To identify the radii of interest, we use data from the extended campaign to analyze accuracy at increasing radii of reporting across different time intervals. Figure~\ref{fig:radius_sweep} shows the ``combined'' accuracy, which emulates a unified ecosystem where devices from Apple, Samsung, and Tile report tag locations independent of proprietary ecosystems. For short time intervals (1 and 10 minutes), accuracy increases with radius until plateauing at approximately 100 meters. For longer time intervals, accuracy shows little improvement beyond 30 meters. Accordingly, we select radii of 10, 30, and 100 meters as we analyze accuracy observed in the extended measurement campaign (involving three tags).

Figure~\ref{fig:accuracy_time_sweeps} (a,b,c) shows that relaxing the responsiveness, i.e., allowing more time to locate a tag within a radius, improves tag accuracy; e.g., the combined tag's accuracy for larger radii (30 and 100 meters) grows from 10\% to 80\% as the responsiveness grows from one to 120 minutes. This observation also applies to a small radius (10 meters, see Figure~\ref{fig:accuracy_time_sweeps}-a) although with a few important differences. First, one minute is too fast to locate a tag within such a small radius, e.g., an accuracy of 8\% versus 15-19\% at larger radii. Second, as we relax the responsiveness, the tag's accuracy increases much slower than what is observed for larger radii, e.g., 60\% versus 70-74\% assuming a responsiveness of 20 minutes. This happens because, as reporting users might move, it is more challenging to correctly report the right location with such a small radius and high responsiveness. Finally, the maximum accuracy caps at 75\%, when considering all tags combined, or 5-7\% less than what is observed for larger radii. Given the slow responsiveness allowed, this reflects errors introduced by approximating a tag's location with the reporting device location, which is unlikely more than 50 meters (see Figure~\ref{fig:rssi}).

If we focus on each tag independently, Figure~\ref{fig:accuracy_time_sweeps}-a shows that SmartTag (blue lines) slightly outperforms AirTag (green lines) at a radius of 10 meters. However, at larger radii, this trend does not hold, with both tags performing similarly at radii of 30 and 100 meters. On the other hand, Tile (orange lines) performs significantly worse than the other two tags, peaking at just 25-33\% on average across radii. These variations are likely influenced by the differing availability of companion devices in the regions visited by the vantage points, which we discuss further in Section~\ref{sec:accuracy:global}. 

\subsection{Impact of COVID-19}
\label{sec:accuracy:covid}
As discussed in Section~\ref{sec:data_collection:in_the_wild}, 
the COVID-19 pandemic and its associated social distancing measures significantly reduced public interactions, likely impacting the availability of companion devices and, consequently, the results in the original study (conducted in mid-2022). To evaluate post-pandemic shifts in tag accuracy, we compare the results of the original and the extended campaign (see Table~\ref{tab:campaigns}) across various mobility and temporal characteristics. This analysis assumes a responsiveness interval of 10 minutes and radii of 10, 50, and 100 meters, as %
in the original study. Figure~\ref{fig:mobility_and_tod} presents our results across three factors: the vantage point's speed, time of day, and day of the week. 
Each bar represents the results from the extended campaign, with shaded regions overlaying the bars to indicate the average accuracy reported in the original campaign. The differences between the bar lengths and the shaded regions highlight changes in tag accuracy following the relaxation of pandemic-related regulations. Note that Tile (orange bars) lack shaded regions as %
it was not analyzed in~\cite{tags_v1}.

\vspace{0.05in}
\subsubsection{Mobility}

\begin{figure*}[t]
\centering
\includegraphics[width=\textwidth]{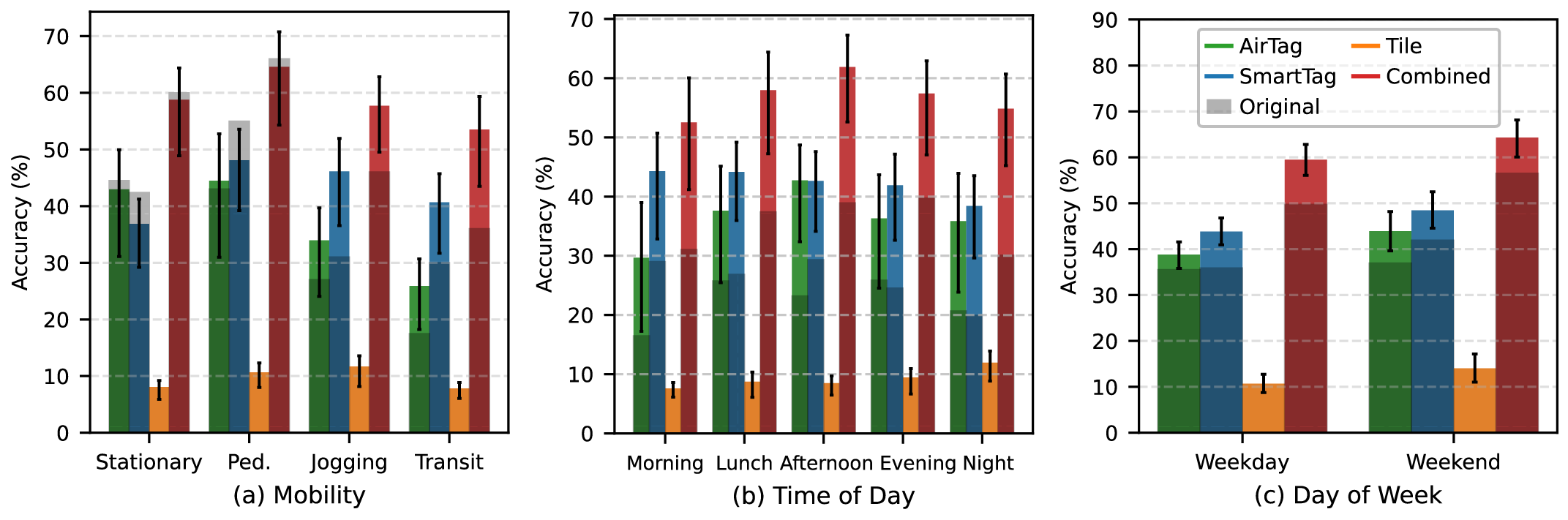}
\caption{\small Evaluation of AirTag, SmartTag, and ``combined'' accuracy in a 10-minute bucket as a function of mobility (a), time of day (b), and day of the week (c). Shaded regions indicate average results obtained in the previous study.}
\label{fig:mobility_and_tod}
\end{figure*}

Figure~\ref{fig:mobility_and_tod}-a shows average tag accuracy—95\% confidence intervals reported as error bars across the different radii considered—as we vary how fast a tag is moving (as per our ground truth). We find that while walking at a pedestrian speed ($<$ 6.0 km/h), the accuracy is maximized for AirTags, SmartTags, and even when combined. The rationale behind this finding is that walking represents a good equilibrium between the number of devices the tag may be exposed to, e.g., higher than when being stationary, and the length of the time window for the Bluetooth signal to be picked up by a location-reporting device. As the speed increases, e.g., when jogging (speed comprised between 6.0 and 12.0 km/h) or in transit ($\geq$ 12.0 km/h), the accuracy deteriorates due to the little time allowed for Bluetooth communication.

Figure~\ref{fig:mobility_and_tod}-a also highlights a %
discrepancy in tag accuracy compared to the prior study. In~\cite{tags_v1}, tag accuracy dropped sharply during jogging (46.1\%) and transit (36.7\%), approximately 20\% and 30\% lower than accuracy at pedestrian speed. Post-pandemic measurements, however, showed improved accuracy of 57.7\% during jogging and 53.5\% in transit, representing only a 7\% and 11\% decrease compared to pedestrian speed. While an overall increase in accuracy was expected with the relaxation of public activity restrictions, this improvement was primarily observed at higher mobility speeds. We conjecture that heightened public activity may increase the likelihood of encountering companion devices moving at similar speeds, such as in crowded public transportation settings. This alignment could help mitigate the spatial errors of location reports typically observed in higher mobility.

\subsubsection{Time of Day / Day of Week}
Figure~\ref{fig:mobility_and_tod}-b shows the average tag's accuracy during different times of the day.
Combined accuracy peaks in the afternoon (2 P.M. to 6 P.M.) at an average of 61.8\%, followed by the lunch period (10 A.M. to 2 P.M.) and evening hours (6 P.M. to 10 P.M.), both averaging 57\%. Lower accuracy is observed during the morning (6 A.M. to 10 A.M.) and night (10 P.M. to 2 A.M.), averaging 52.5\% and 54.8\%, respectively, likely due to reduced public activity during these periods. We next explore the potential impact of weekdays and weekends on the accuracy. Figure~\ref{fig:mobility_and_tod}-c highlights an increase of 4.8\% in combined accuracy on weekends compared to weekdays, likely driven by higher outdoor activity levels. Additionally, both plots indicate improved accuracy across various times of the day and days of the week compared to the findings in~\cite{tags_v1}. This further suggests that social-distancing protocols and reduced public interactions during the pandemic had a notable impact on tag performance.

\subsection{A Global View on Tag Accuracy.}
\label{sec:accuracy:global}
We leverage the wide geographic distribution of our vantage points in the extended campaign, spanning 123 cities, to explore how the varying availability of companion devices impacts tag accuracy across regions. To control for potential skews from analyzing sparsely-populated areas, we incorporate the latest (November 2023) population density data from Kontur~\cite{kontur_population_dataset}, which provides estimates based on satellite imagery of building density within H3 hexagons at resolution 8. Using this dataset, we classify the vantage points' locations into three levels of urbanization: urban ($\geq$1,500 inhabitants per km$^2$), suburban (300-1,500 inhabitants per km$^2$) and rural ($<$ 300 inhabitants per km$^2$), following the widely used classification method established by EuroStat~\cite{eurostat_2011_degurbaclassification}. This reveals that 81.7\% of our data points were located in urban areas, leading us to exclude suburban and rural data for consistency in our subsequent analysis.

\vspace{0.05in}
\subsubsection{AirTag and SmartTag}
We first evaluate the influence of companion device density on the accuracy of AirTags and SmartTags, respectively. While precise information on Apple and Samsung device popularity is challenging to collect in the wild, we approximate it at the national level using smartphone market share (MS) statistics obtained from StatCounter Global Stats~\cite{statcounter_mobile_market_share}. Based on this dataset, we categorize countries into high ($>$40\%), medium (20–40\%), and low ($<$20\%) market share groups for each manufacturer. 
See Table~\ref{tab:countries} in Appendix~\ref{appendix:marketshare} for a summary of urbanization levels and smartphone market share data. 

\begin{figure*}
    \centering
    \includegraphics[width=\textwidth]{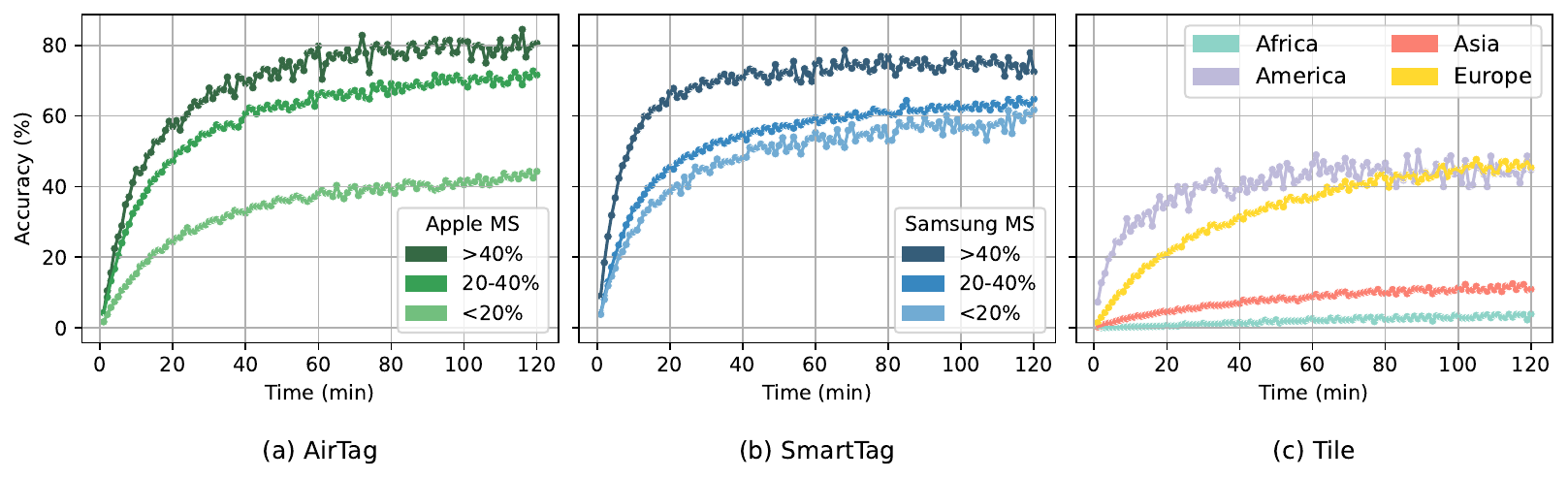} 
    \caption{Accuracy of Tags at a 10-meter radius. (a) AirTag accuracy by Apple's mobile market-share (MS). (b) SmartTag accuracy by Samsung's mobile MS. (c) Tile accuracy across continents.} 
    \label{fig:accuracy_time_sweeps_vendor}    
\end{figure*}

Figure~\ref{fig:accuracy_time_sweeps_vendor}-(a,b) plots the accuracy of AirTags and SmartTags across countries grouped by the MS of their respective companion devices. The analysis uses a 10-meter radius, which is the maximum distance at which AirTags switch to Ultra Wideband (UWB) to provide precise directional instructions for locating the device\footnote{For SmartTag+ and SmartTag2, the maximum UWB range is 15 meters. Tile does not support UWB at the time of writing.}. 
For both AirTags and SmartTags, accuracy correlates with the regional popularity of companion devices. AirTags plateau at 79.8\% accuracy after 60 minutes in 5 countries with high iPhone MS, compared to 66\% in 13 countries with medium MS and 38\% in 11 countries with low MS. Similarly, SmartTags reach 70.8\% accuracy within 40 minutes in 3 countries with high Samsung smartphone popularity, while achieving 54.8\% in 20 countries with medium MS and 48.0\% in 6 countries with low MS. 

\vspace{0.05in}
\subsubsection{Tile} 
Tile's accuracy is significantly lower than that of AirTag and SmartTag across our vantage points, despite its compatibility with both iOS and Android.  While we could not get reliable data on regional popularity of Life360/Tile, Figure~\ref{fig:accuracy_time_sweeps_vendor}-c shows that Tile's accuracy varies notably across continents. The highest accuracy is observed in the Americas, plateauing at %
44.3\% after 50 minutes, followed by Europe, where accuracy steadily increases over time, reaching a similar level after 95 minutes. In contrast, Tile performs poorly in Asia and Africa, with maximum accuracies of only 12.5\% and 3.9\%, respectively. This low accuracy can be likely attributed to limited adoption of Life360/Tile. Additionally,  %
Tile's best performance, observed in the Americas (Figure~\ref{fig:accuracy_time_sweeps_vendor}-c, violet line), initially surpasses AirTag's accuracy in regions with low Apple device popularity (Figure~\ref{fig:accuracy_time_sweeps_vendor}-a, light-green line)  However, both converge to approximately 40\% accuracy after 60 minutes. On the other hand, Tile consistently lags behind SmartTag's accuracy after 10 minutes, even in regions with low Samsung device density (Figure~\ref{fig:accuracy_time_sweeps_vendor}-b, light-blue line).

\begin{figure*}[t]
\centering
\includegraphics[width=\textwidth]{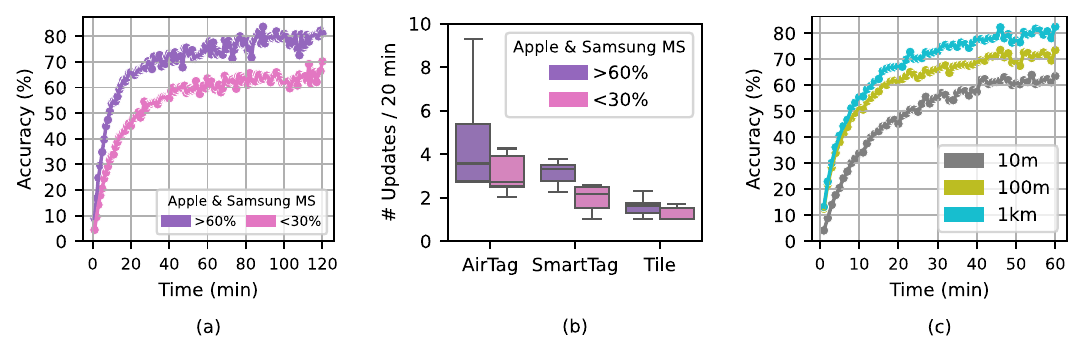}
\caption{Comparison of (a) combined accuracy and (b) average number of location updates per 20-minute window in countries with high MS ($>$60\%) vs. low MS ($<$30\%) of Apple and Samsung mobile devices. (c) shows the combined accuracy as a function of radius in countries with low Apple/Samsung MS.}
\label{fig:low_market_analysis}
\end{figure*}

\vspace{0.05in}
\subsubsection{Discussion}
The previous results suggest that a tag accuracy is significantly influenced by the varying prevalence of its companion devices across geographical regions. To further examine this hypothesis, we compare the performance between two groups of countries, categorized by their Apple and Samsung mobile MS. The first ``high-MS'' group comprises five countries where Apple and Samsung jointly hold over 60\% MS, with each company maintaining at least 30\%. The second ``low-MS''group consists of six countries where the two companies both have low presence in the mobile device market, with their combined MS below 30\%.  

Figure~\ref{fig:low_market_analysis}-a shows that tags deployed in the high-MS group (purple) consistently outperformed those in the low-MS group (pink), averaging 15.2\% higher combined accuracy after 10 minutes. This performance disparity can be attributed to differences in location update rates, as tags in regions with greater densities of companion devices are expected to benefit from more frequent location updates (see section~\ref{sec:exp:update_freq}). Figure~\ref{fig:low_market_analysis}-b quantifies this relationship, showing the average number of location updates per tag type over 20-minute intervals.  In high-MS countries, AirTags and SmartTags averaged 3.57 and 3.31 updates, respectively. In contrast, these values dropped to 2.73 and 2.17 in low-MS countries, suggesting that tags in these regions were more likely to suffer from higher temporal errors.

These temporal errors are further compounded by spatial errors, as manufacturers approximate tag locations by aggregating reports from multiple companion devices within fixed time windows. In low-MS countries, the sparse data sources may also result in imprecise geolocation calculations, with updated locations being non-negligibly distant from the ground truth.  
Figure~\ref{fig:low_market_analysis}-c shows the combined accuracy in low-MS countries under increasingly relaxed geographical precision criteria, where location reports are classified as ``hits'' if they fall within larger radii from the actual GPS location. At a 100-meter radius, 61.2\% of updates were accurate after 20 minutes (when at least one location update can be expected for any tag types, per Figure~\ref{fig:low_market_analysis}-b), a 16.1\% improvement over the 10-meter radius accuracy of 45.1\%. Expanding the radius to 1 km increased accuracy to 67.2\%. Together, these findings highlight the challenges of using tags with limited companion device density, where infrequent updates and imprecise location calculation lead to reduced overall accuracy.

\vspace{0.05in}
\subsubsection{Limitation}
Our dataset is collected from few volunteers per country, which may not fully represent the general experience of tag owners in those countries. Although volunteers were instructed to roam public areas as extensively as possible while carrying the tags, the dataset could still be significantly influenced by individuals they interacted with frequently—a factor beyond our control. Consequently, we acknowledge the limitations in generalizing these findings to draw conclusive assertions on geographical characteristics of tags' performance. Nonetheless, our results report a consistent trend across countries when grouped by the popularity of companion devices. This underscores the tags' dependence to their respective ecosystems for timely and accurate location updates, reflecting a systemic limitation that extends beyond an individual's control.   

\section{Report Reliability}
\label{sec:reliability}
Reliability refers to the likelihood that a tag's location report accurately reflects its current ground truth, addressing a key question in real-world use: \textit{to what extent can users trust a freshly updated tag location}? Complementing the broader accuracy evaluation, reliability provides a more immediate measure of confidence in \textit{individual} location updates. Section~\ref{sec:reliability:methods} explains the three metrics employed for quantifying reliability, which is followed by the discussion of results from the extended campaign in Section~\ref{sec:reliability:res}. 

\subsection{Evaluation Metric}
\label{sec:reliability:methods}
\noindent We quantify tag reliability using the following three metrics:
\vspace{0.05in}
\subsubsection{Confidence Interval of $\sigma$}
The confidence interval of $\sigma$, which we denote as CI$_{\sigma}$ for brevity, quantifies the probability that a tag's reported position lies within the specified radius ($\sigma$) from the true location (refer to Section~\ref{sec:method:description}). 
This reflects how accurately tags estimate their distance from the ground truth in each of their location updates. In real-world scenarios, where location tracking often relies on a single update (e.g., locating a lost item), CI$_{\sigma}$ represents the certainty of the reported information, making it a practical measure of reliability. Since none of the tag manufacturers we study disclose this metric, we empirically determine CI$_{\sigma}$ for each tag type. %

To calculate CI$_{\sigma}$, we match each tag's reported location to the most recent GPS data recorded within the same one-minute interval. For each valid match, we record the following information: the latitude, longitude, and radius ($\sigma_{\text{tag}}$) reported by the tag; the corresponding GPS latitude, longitude, and radius ($\sigma_{\text{GPS}}$); and the spatial error ($\delta$), which is the distance between the tag's reported location and the GPS location.

Using these matched data points, we derive CI$_{\sigma}$ %
for each tag type by  calculating the percentage of matches where the spatial error ($\delta$) is less than the tag's reported radius ($\sigma_{\text{tag}}$). Intuitively, this is equivalent to the percentage of GPS coordinates that fall within the circle centered at the tag's coordinates with a radius of $\sigma_{\text{tag}}$. This approach provides a straightforward measure of reliability: a higher confidence interval indicates that the ground truth location more frequently falls within the tag's estimated margin of error.

\begin{figure}
    \centering    
    \includegraphics[width=\linewidth]{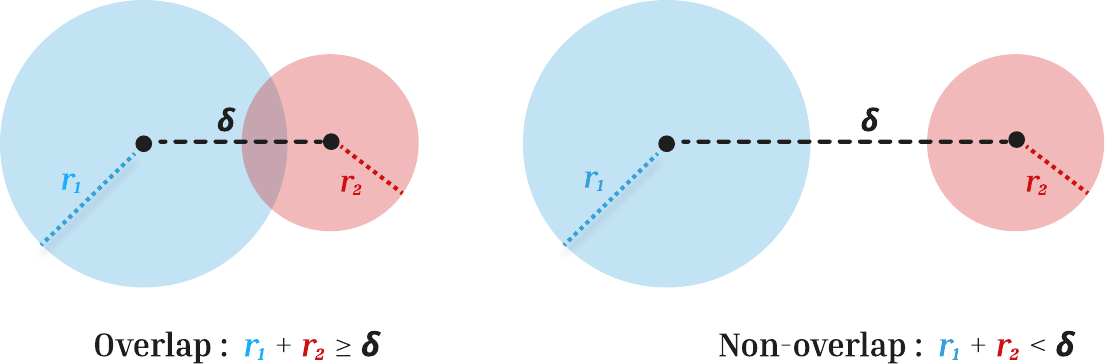} 
    \caption{Examples of overlapping circles (left) and non-overlapping circles (right), based on the sum of their radii relative to the distance between their centers, or $\delta$.} 
    \label{fig:overlap_circles}    
\end{figure}

\subsubsection{Fit to Rayleigh Distribution} %
We test a common assumption in location systems: that $\delta$ follows a Rayleigh distribution parameterized by $\sigma$ -- i.e., $\delta \sim \text{Rayleigh}(\sigma)$~\cite{zandbergen2008positional}.  
This assumption relies on the premise that latitude and longitude errors are independent and normally distributed. If this assumption holds for BLE tags, users can leverage the reported $\sigma_{\text{tag}}$ value in each location update to estimate the probability of the ground truth being within any specified distance (e.g., the likelihood of the ground truth being within X meters).

To evaluate the Rayleigh distribution's goodness of fit to the tags dataset, we conduct a Monte Carlo simulation~\cite{mooney1997monte}. For each matched data point between a tag and GPS measurement, we generate two sets of 10,000 samples: (1) radii drawn from \( r_{\text{1}} \sim \text{Rayleigh}(0.68*\sigma_{\text{tag}})  \), and (2) radii drawn from \( r_{\text{2}} \sim \text{Rayleigh}(0.68 * \sigma_{\text{GPS}}) \),  where 0.68 reflects the expected confidence interval for $\sigma$ in location systems. These sets represent the estimated margins of spatial error for the tag and GPS measurements, assuming their $\sigma$ values conform to the standard confidence interval.

The fit of each matched data point is assessed by calculating the probability of overlap between pairs of sampled radii from the two distributions. As illustrated in Figure~\ref{fig:overlap_circles}, overlap is defined as the condition where the sum of a radii pair exceeds $\delta$. The average overlap probability for each data point thus quantifies how well the tag's location update aligns with GPS measurements when their respective margins of error are considered under the Rayleigh distribution assumption.
We determine the overall fitness for each tag type by averaging the overlap probabilities across all their valid matches. If the overall fitness is high for a particular tag type, it suggests that the Rayleigh distribution provides a strong model for its spatial errors, enabling users to make probabilistic estimates of the tag's true location based on the reported $\sigma_{\text{tag}}$ value.

\vspace{0.05in}
\subsubsection{Spatial Error Distribution} The statistical distribution of spatial error $\delta$ is another useful metric for evaluating reliability, as it models tag's location reports in relation to the true position. To explore this, we test whether common statistical distributions—including beta, gamma, logistic, cosine, log-normal, and skew-normal—can effectively model the spatial error distribution for each tag type. Parameters for each candidate distribution were estimated using maximum likelihood estimation (MLE) based on empirical position errors. We select the best-fit distribution for each tag type via the Kolmogorov–Smirnov (KS) test~\cite{scipy_kstest}, and additional measures such as skewness and kurtosis were analyzed to provide deeper insights into the distributions' characteristics.

\subsection{Results}
\label{sec:reliability:res}
\subsubsection{Confidence Interval of $\sigma$}

Table~\ref{tab:sigma_ci} details the empirical CI$_{\sigma}$ for each tag type. Out of 5,186 matched data points, AirTags achieve a 68.7\% CI$_{\sigma}$, comparable to Android GPS, which reports a 68\% CI$_{\sigma}$ according to its documentation~\cite{android_location_request}. In contrast, Tile tags achieve a lower CI$_{\sigma}$ of 55.9\% based on 764 matches. SmartTags exhibit the lowest reliability, with only 29.8\% of 4,573 matched data points falling within the reported radius $\sigma$, suggesting that their location reports underestimate the distance from the ground truth. 

Table~\ref{tab:sigma_ci} further shows that while AirTag and SmartTag demonstrate similar ranges of observed spatial error ($\delta$), they differ significantly in their accuracy of estimating the distances from ground truth. SmartTags tend to underestimate, with a median $\sigma$ of 11.7 meters compared to a median observed spatial error of 22.82 meters. In contrast, AirTags make more broad estimates, with a median $\sigma$ of 36.15 meters, while the median observed spatial error is lower at 22.02 meters. Such loose-fitting approach likely contributes to AirTag's higher CI$_{\sigma}$ compared to other tags.

\begin{figure}[h]
    \includegraphics[width=\linewidth]{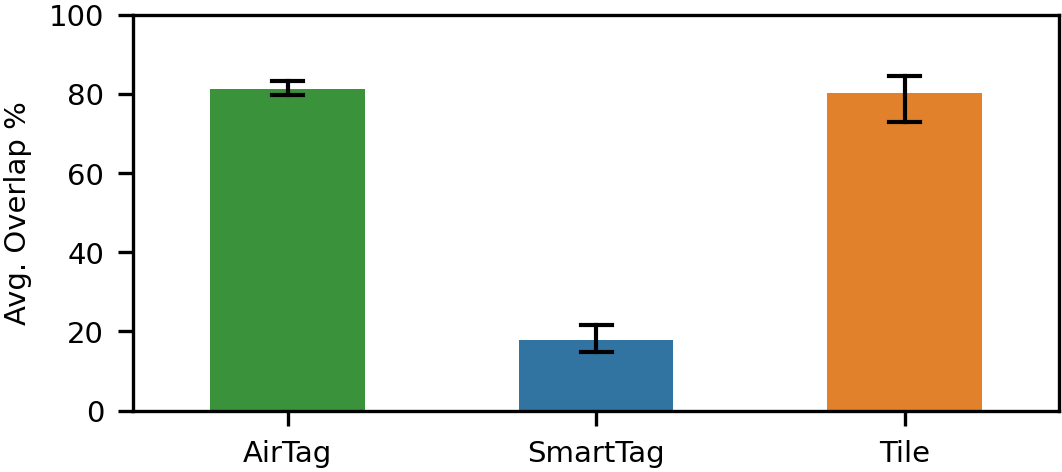} 
    \caption{Median of average overlap probability across tag types. Overlap probability is calculated as the percentage of radii pairs sampled from Rayleigh distributions (parameterized by $\sigma$ from the tag and corresponding GPS measurement) whose sum exceeds the distance between the tag and GPS measurement. The black error line indicates 95\% CI.} 
    \label{fig:rayleigh_overlap}    
\end{figure}

\begin{table}
\centering

\begin{tabular}{|p{0.9cm}|p{1cm}|p{2.15cm}|p{2.25cm}|p{0.5cm}|}
\hline
\textbf{Tag} & \textbf{\# Match} & \textbf{$\delta$ (m)} & \textbf{$\sigma$ (m)} & \textbf{{$\text{CI}_\sigma$}} \\
\hline
AirTag & 5,186 & 22.02 (10.0-48.56) & 36.15 (29.33-43.16) & 68.7\% \\
\hline 
SmartTag & 4,573 & 22.82 (11.38-45.8) & 11.7 (6.34-19.1) & 29.8\% \\
\hline 
Tile & 764 & 17.28 (6.60-34.68) & 19.7 (7.92-35.0) & 55.9\% \\
\hline 
\end{tabular}

\caption{Summary of matched data points across tag types. For observed spatial error ($\delta$) and estimated spatial error ($\sigma$), the median and IQR are reported. {$\text{CI}_\sigma$} is the percentage of location reports where $\sigma$ $\geq$ $\delta$.}
\label{tab:sigma_ci}
\vspace{-0.2in}
\end{table}

\subsubsection{Fit to Rayleigh Distribution}
Figure~\ref{fig:rayleigh_overlap} presents the median of the average overlap probability across all location updates for each tag type. The results show that AirTags and Tile achieve median average overlap probabilities of 81.2\% and 80.3\%, respectively. This indicates a reasonable fit between the observed errors and the Rayleigh distribution modeled on the reported $\sigma_{\text{tag}}$, demonstrating their potential to provide probabilistic estimates of the ground truth location at any given distance.  In contrast, SmartTag exhibit lower average probabilities of 17.8\% in median, underscoring the Rayleigh distribution's limitations in modeling its spatial errors.

\begin{figure*}
    \includegraphics[width=\textwidth]{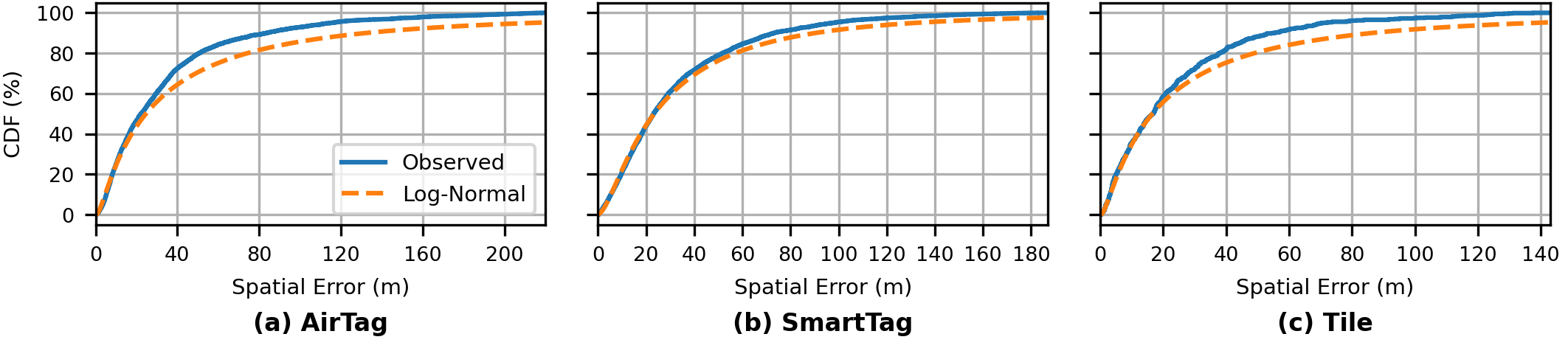} 
    \caption{CDF of observed spatial error compared to the best-fit log-normal distribution for each tag type.} 
    \label{fig:error_model_cdf}    
\end{figure*}
\subsubsection{Spatial Error Distribution}
Our analysis of 13 candidate distributions reveal that the spatial errors for all three tags are best modeled by a log-normal distribution. Figure~\ref{fig:error_model_cdf} compares the cumulative distribution function (CDF) of the observed spatial error with the fitted log-normal model. For AirTag, the log-normal model closely aligns with the observed spatial error for approximately 50\% of location updates, where the spatial error was 24.4 meters, before slightly underestimating the error compared to the empirical data. A similar pattern is observed for the other two tags, with the tailored log-normal models closely matching the empirical spatial error for over 60\% of data points. These results suggest that the log-normal distribution provides a suitable statistical model for spatial errors in BLE tags.

Table~\ref{tab:lognorm_stats} summarizes the statistics of the log-normal distributions for the three tag types. The results show that all three distributions are asymmetric (skewness $>$ 0) and leptokurtic (kurtosis $>$ 3), indicating heavier tails and sharper peaks in their spatial error distributions. These characteristics suggest that the devices occasionally produce location readings with significant deviations from the ground truth, as reflected in their high standard deviations, which ranged from 56.76 (SmartTag) up to 124.87 (AirTag) meters.

\begin{table}
\centering
\begin{center}
\begin{tabular}{|l|c|c|c|c|c|}
\hline
\textbf{Device} & \textbf{Mean} & \textbf{Std.} & \textbf{Skewness} & \textbf{Kurtosis}  & \textbf{p-value}\\ \hline
AirTag   & 58.27 & 124.87 & 16.33 & 1430.53 &  0.016 \\ \hline
SmartTag & 40.39 & 56.76  & 6.65  & 133.30 & 0.01 \\  \hline
Tile     & 37.93 & 78.1  & 14.87 & 1111.08 & 0.04 \\ \hline
\end{tabular}

\end{center}
\caption{Statistics of the log-normal distribution models that best fit the empirical spatial error for each tag type.}
\label{tab:lognorm_stats}
\vspace{-0.2in}
\end{table}

\vspace{0.05in}
\subsubsection{Takeaway}
These findings highlight critical differences in the reliability of location tracking among tag types, with implications for their practical use. Specifically, higher reliability minimizes the necessity for extensive calibration, i.e., allowing users to trust the reported locations directly without additional processing or filtering.  AirTags exemplify this high reliability as they maintain a relatively consistent relationship between reported $\sigma$ and actual spatial errors, performing on par with conventional GPS systems.   Tile, while less reliable than AirTags, provides a moderate level of reliability,  which could be sufficient for low-risk, everyday uses without requiring supplementary adjustments. 

In contrast, recognizing that SmartTag tends to heavily underestimate $\sigma$ offers a cautionary note for applications that demand precise margin of positional error. Without additional  processing, filtering, or more conservative approach for estimating error, relying solely on a SmartTag's reported $\sigma$ could result in critical misjudgments about the device's true location. To mitigate this, Samsung and platform developers should consider applying corrective factors, such as expanding the reported error radius, or implementing machine-learning-based filters that learn from historical discrepancies. Additionally, developers of location-based systems could incorporate adaptive selection mechanisms, where the system automatically chooses the most reliable tag or a combination of tags for a given context.

\section{Conclusion}
This paper evaluates the performance of three location tags---AirTag, SmartTag, and Tile---through both controlled experiments and real-world scenarios. Drawing on data collected in 29 countries, the study examines how factors such as user movement, companion device availability, and regional differences impact the usability of these tracking devices. The findings emphasize that the density of nearby compatible devices is a critical factor in determining their performance. Comparing our results to a prior study~\cite{tags_v1}, we show that increased public mobility after pandemic restrictions may have benefited the ``in-the-wild'' accuracy of AirTag and SmartTag. In particular, AirTag showed notable progress in location update frequency, attributed to recent software updates. Tile, on the other hand, seems to struggle due to relatively limited adoption on a global scale. We discover a tendency for SmartTag to underestimate its margin of error from the ground truth, raising questions over its reliability. The empirical spatial errors across tag devices conform to a lognormal distribution, which suggests the need for refined methods to better handle occasional large inaccuracies in location reporting. 

\section{Ethics}
\label{sec:ethics}
We obtained IRB approval (HRPP-2021-185) and informed participants of our data collection practices through a consent form. While we collect GPS data, we do not gather any identifiable or sensitive personal information. 
\newpage

\balance
\bibliographystyle{ieeetr}  
\bibliography{literature}
\newpage
\begin{appendices}

\section{Breakdown of Measurement Campaigns}
\label{appendix:campaign_breakdown}
Tables~\ref{tab:exp_breakdown_orignal} and ~\ref{tab:exp_breakdown_extended} provide an overview of two in-the-wild experiments: the original measurement campaign across 5 countries (March to August 2022)~\cite{tags_v1}, and the extended measurement campaign across 29 countries (December 2023 to January 2024), respectively. For each country, we detail the number of different cities, volunteers, measurement duration (in days), and total distance covered.

\begin{table}[hb!]
\begin{center}
\begin{tabular}{|c|c|c|c|c|}
\hline
Country & \# Cities & \# Volunteers & \# Days & Distance (Km) \\
\hline
US & 2 & 1 & 30 & 907 \\
IT & 10 & 1 & 28 & 3,395 \\
AE & 2 & 1 & 52 & 3,680 \\
PK & 1 & 1 & 2 & 194 \\
CH & 1 & 1 & 3 & 92 \\
DE & 3 & 1 & 5 & 1,112 \\
\hline
\end{tabular}

\end{center}
\caption{Summary of in-the-wild experiments conducted during the original measurement}
\label{tab:exp_breakdown_orignal}
\end{table}

\begin{table}[hb!]
\begin{center}
\begin{tabular}{|c|c|c|c|c|}
\hline
Country & \# Cities & \# Volunteers & \# Days & Distance (Km) \\
\hline
AT & 1 & 1 & 2 & 197.1354 \\
BD & 4 & 2 & 15 & 967.0203 \\
CN & 8 & 4 & 18 & 2706.5775 \\
CZ & 7 & 2 & 30 & 1364.9414 \\
DE & 4 & 2 & 20 & 4912.9283 \\
EC & 5 & 1 & 29 & 1813.2888 \\
EG & 12 & 2 & 13 & 2225.0126 \\
ET & 1 & 1 & 11 & 2231.6615 \\
FR & 11 & 3 & 19 & 3037.4548 \\
GB & 38 & 3 & 29 & 34053.9036 \\
GH & 14 & 1 & 12 & 37981.5489 \\
HK & 2 & 1 & 1 & 113.0431 \\
IN & 32 & 4 & 17 & 26297.3406 \\
IT & 6 & 2 & 20 & 4104.2046 \\
JM & 17 & 1 & 29 & 13330.904 \\
KE & 6 & 1 & 12 & 1930.7876 \\
KR & 7 & 1 & 15 & 2749.7775 \\
LB & 4 & 1 & 16 & 4502.397 \\
MD & 4 & 1 & 12 & 1060.4525 \\
MY & 5 & 1 & 13 & 1216.2438 \\
NL & 1 & 3 & 3 & 428.9035 \\
NP & 6 & 2 & 37 & 3437.1752 \\
OM & 8 & 5 & 6 & 11166.8492 \\
PH & 26 & 1 & 30 & 2980.9204 \\
PK & 8 & 3 & 15 & 5759.8113 \\
QA & 6 & 4 & 11 & 2849.2939 \\
TH & 13 & 1 & 10 & 874.6035 \\
TZ & 4 & 2 & 12 & 1755.5241 \\
UZ & 11 & 1 & 10 & 1571.2417 \\
\hline
\end{tabular}

\end{center}
\caption{Summary of in-the-wild experiments conducted during the extended measurement}
\label{tab:exp_breakdown_extended}
\end{table}

\section{Smartphone Market-share}
\label{appendix:marketshare}
Using Kontur's November 2023 population density data~\cite{kontur_population_dataset}, we filtered for urban locations in our dataset, following EuroStat's urbanization classification ($\geq$1,500 inhabitants per km$^2$) ~\cite{eurostat_2011_degurbaclassification}. 
We utilized smartphone market share data from StatCounter Global Stats~\cite{statcounter_mobile_market_share} as a proxy for Apple and Samsung device prevalence.
Table~\ref{tab:countries} presents the results across 29 countries visited during the extended measurement campaign, showing the percentage of data points collected in urban areas and the respective market shares for both Apple and Samsung mobile devices.

\begin{table}[hb!]
\begin{center}
\begin{tabular}{|c|c|c|c|}
\hline
Country & Urban (\%) & Apple (\%) & Samsung (\%) \\
\hline
AT & 99.956498 & 42.33 & 34.42 \\
BD & 85.433607 & 4.72 & 20.05 \\
CN & 89.331057 & 22.37 & 1.39 \\
CZ & 100 & 33.62 & 24.02 \\
DE & 81.569309 & 38.67 & 34.23 \\
EC & 64.984699 & 16.77 & 31.74 \\
EG & 69.345351 & 12.95 & 25.83 \\
ET & 100 & 4.62 & 47.51 \\
FR & 69.297343 & 32.78 & 31.93 \\
GB & 91.472025 & 52.46 & 29.54 \\
GH & 99.600776 & 17.16 & 22.97 \\
HK & 2.018822 & 50.00 & 28.49 \\
IN & 88.748496 & 4.02 & 13.65 \\
IT & 57.399770 & 31.50 & 29.14 \\
JM & 74.990379 & 32.61 & 52.89 \\
KE & 60.791558 & 2.39 & 19.41 \\
KR & 94.232732 & 26.53 & 68.74 \\
LB & 96.190833 & 28.84 & 36.51 \\
MD & 98.750951 & 25.09 & 32.69 \\
MY & 100 & 29.93 & 14.75 \\
NL & 99.145458 & 43.39 & 36.17 \\
NP & 71.553505 & 12.25 & 24.92 \\
OM & 0 & 25.42 & 24.83 \\
PH & 82.753575 & 12.66 & 13.76 \\
PK & 99.909871 & 4.62 & 15.72 \\
QA & 65.222062 & 21.97 & 23.21 \\
TH & 100 & 33.69 & 20.12 \\
TZ & 52.004633 & 6.61 & 23.63 \\
UZ & 82.136526 & 10.01 & 27.66 \\
\hline
\end{tabular}

\end{center}
\caption{Percentage of data points located in urbanized areas and smartphone market share for Apple and Samsung across countries.}
\label{tab:countries}
\end{table}

\end{appendices}

\end{document}